\documentclass[11pt, letterpaper]{article}

\usepackage[margin=1in]{geometry}
\usepackage{setspace}
\usepackage[noblocks]{authblk} 

\usepackage{tgtermes}
\usepackage{newtxtext}
\usepackage{newtxmath}
\usepackage{bm}
\usepackage{array}

\usepackage{algorithm}
\usepackage{algpseudocode}
\usepackage{tikz}
\usepackage{booktabs}
\usepackage{graphicx}

\usepackage{amsthm}

\usepackage[numbers]{natbib}
 \bibpunct[, ]{(}{)}{,}{a}{}{,}%
\usepackage[colorlinks=true, linkcolor=blue, citecolor=blue, urlcolor=blue]{hyperref}

\title{Beyond AI advice---independent aggregation boosts human--AI accuracy}

\author[1]{Julian Berger}
\author[2]{Pantelis P. Analytis}
\author[3]{Ville Satopää}
\author[1, 4]{Ralf HJM Kurvers}

\affil[1]{Max Planck Institute for Human Development}
\affil[2]{University of Southern Denmark}
\affil[3]{INSEAD}
\affil[4]{Science of Intelligence, Technical University Berlin}

\date{\today}

\begin{document}

\maketitle

\begin{abstract}
Artificial intelligence (AI) is broadly deployed as an advisor to human decision-makers: AI recommends a decision and a human accepts or rejects the advice. This approach, however, has several limitations: People frequently ignore accurate advice and rely too much on inaccurate advice, and their decision-making skills may deteriorate over time. Here, we compare the AI-as-advisor approach to the hybrid confirmation tree (HCT), an alternative strategy that preserves the independence of human and AI judgments. The HCT elicits a human judgment and an AI judgment independently of each other. If they agree, that decision is accepted. If not, a second human breaks the tie. For the comparison, we used 10 datasets from various domains, including medical diagnostics and misinformation discernment, and a subset of four datasets in which AI also explained its decision. The HCT outperformed the AI-as-advisor approach in all datasets. The HCT also performed better in almost all cases in which AI offered an explanation of its judgment. Using signal detection theory to interpret these results, we find that the HCT outperforms the AI-as-advisor approach because people cannot discriminate well enough between correct and incorrect AI advice. Overall, the HCT is a robust, accurate, and transparent alternative to the AI-as-advisor approach, offering a simple mechanism to tap into the wisdom of hybrid crowds.

\end{abstract}

\section*{Introduction}

The performance of AI systems is now on par with, and sometimes even superior to, human performance in high-stakes domains such as medical diagnostics \citep{zoller2025human,hekler2019pathologist, hekler2019superior}, geopolitical forecasting \citep{benjamin2023hybrid, karger2025forecastbench}, and misinformation detection \citep{groh2022deepfake, deverna2024fact}. Yet legal and ethical considerations make it imperative that humans remain in the loop. A human expert, unlike an algorithm, can be held legally and ethically accountable \citep{santoni2021four}, and society broadly expects humans to audit AI outputs to ensure safe and ethical decision-making \citep{rudin2019stop, obermeyer2019dissecting}. Regulatory frameworks such as the European AI Act 
mandate such requirements and explicitly call for human oversight in high-stakes applications. Consequently, it is crucial to not only build better AI systems, but also to establish effective workflows for human--AI interactions.

The dominant paradigm for deploying predictive AI alongside human experts is the AI-as-advisor approach. In this setup, AI provides a recommendation to a human decision-maker, who then makes the final call. While this approach preserves human agency and tends to improve decision-making \citep{vaccaro_when_2024}, it cannot fully capitalize on the complementary potential of human and AI judgment because it is constrained by a procedural weakness: By placing the human decision after the AI’s output, the onus is on the human to adequately distinguish between accurate and inaccurate AI advice. Yet humans struggle to do so, displaying both underreliance (rejecting accurate advice) \citep{buccinca2021trust, bansal2021does} and overreliance (accepting inaccurate advice) \citep{buccinca2021trust, yin2025designing, chiang2023two}. Moreover, the AI-as-advisor approach risks humans defaulting to AI's suggestions, especially when humans are aware of AI's overall superior performance in a particular domain and are under time pressure \citep{parasuraman1997humans, skitka1999does}, potentially leading to the deskilling of human decision-makers \citep{bainbridge1983ironies, budzyn2025endoscopist, berzin2025preserving}. The AI-as-advisor approach thus creates a paradox: It preserves the human capacity to have the final say, but can degrade people's ability to do so effectively.

Here we propose the hybrid confirmation tree (HCT) \citep{andersen2023confirmation, berger2026hybridconfirmationtreerobust}, an alternative heuristic for human--AI decision-making in which an AI system independently assesses cases alongside humans. In the HCT, a human and an AI system make initial decisions independently. If they agree, the choice is accepted; if they disagree, a second human is consulted to break the tie. The HCT modifies the interaction structure such that a human approves the final choice, and human decision-making is outside AI influence. By preserving independence, the HCT can reap the benefits of combining independent judgments \citep{de2014essai, grofman1983thirteen, kurvers2019detect, herzog_ecological_2019} while evading the risk of premature consensus and herding that can plague sequential or socially influenced decision-making processes \citep{frey2021social, lorenz2011social, jia2023herding}. 

We compared the HCT to the AI-as-advisor approach across 10 datasets---containing a total of over 41,000 human decisions---from the domains of medical diagnostics, misinformation detection, sentiment classification, deception detection, and rearrest predictions. The HCT outperformed the AI-as-advisor approach across all datasets. 
Even when humans had access to explainable AI advice---in an additional comparison based on more than 50,000 human choices---the HCT performed better, except when human accuracy was at the level of a coin flip. 
Therefore, in almost all practical applications involving trained decision-makers (e.g., dermatologists), the HCT is superior to the AI-as-advisor approach. The HCT takes advantage of correct AI decisions more often than humans who directly receive correct AI advice: In cases of initial disagreement, the tiebreaker in the HCT frequently agrees with the AI's correct choice, whereas people who receive correct AI advice but disagree with it do not adopt it often enough. 

To better understand the performance edge the HCT has over the AI-as-advisor approach, we built a signal detection model that describes the difference in their performance as a function of human decision-makers' ability to distinguish correct from incorrect AI advice and their general propensity to rely on advice. This framework describes our empirical results accurately, even explaining between-dataset differences, and reveals two complementary reasons that the AI-as-advisor approach falls short: Humans cannot discriminate well enough between correct and incorrect AI advice and, overall, they do not rely on AI advice enough.

\begin{figure*}[!h]
    \centering
    \includegraphics[width = \textwidth]{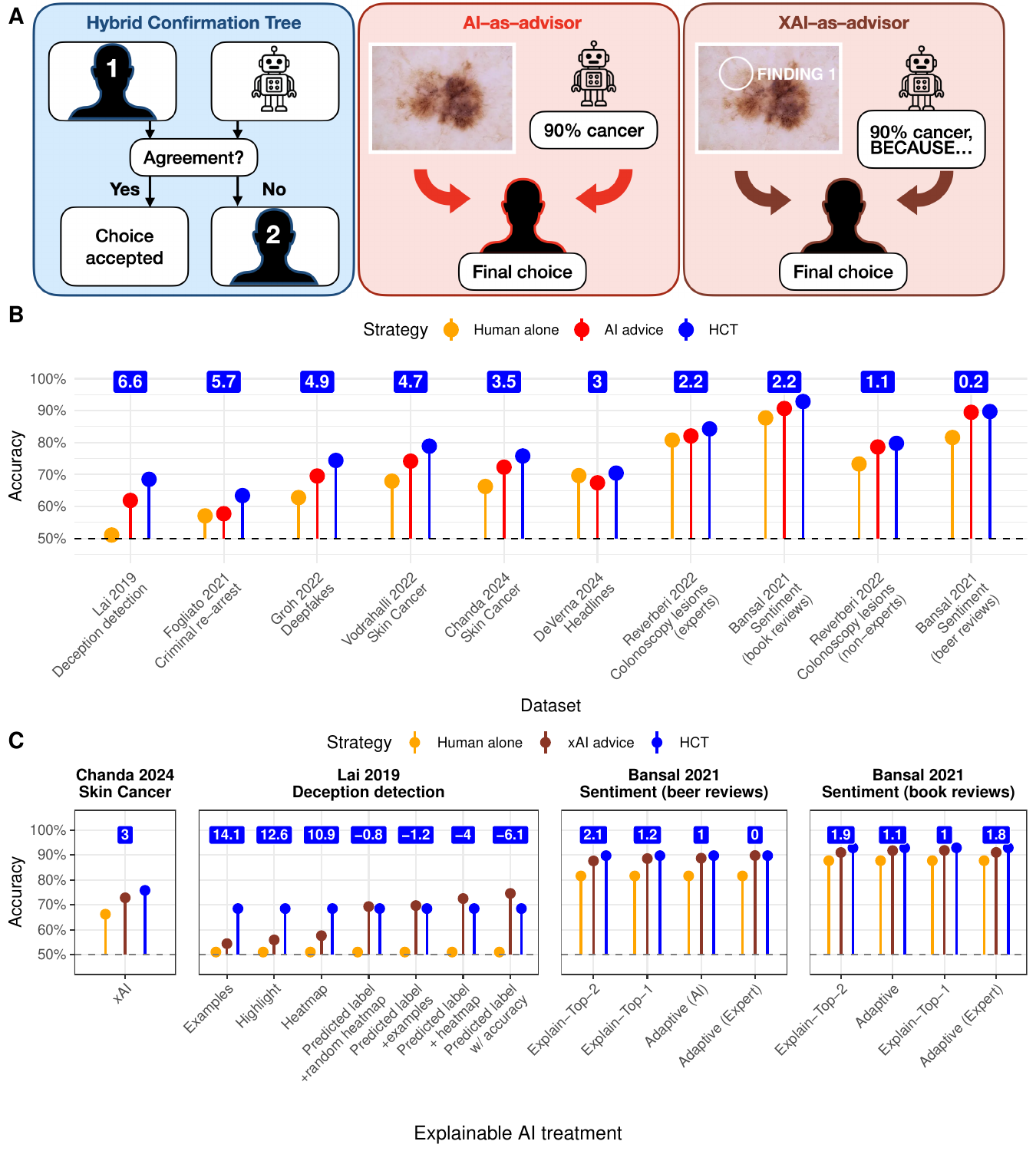}
    \caption{Workflow and performance of the hybrid confirmation tree (HCT) and the AI-as-advisor approach. (A) Workflow of the HCT and the AI-as-advisor approach with and without AI explanations (XAI-as-advisor). (B) Mean accuracy of the HCT, the AI-as-advisor approach, and humans without AI advice across datasets. (C) Mean accuracy of the HCT, the XAI-as-advisor approach, and humans without AI advice across datasets and experimental conditions of the XAI-as-advisor. Blue boxes show the accuracy improvement of the HCT over the AI-as-advisor. Results in (B) are ranked based on this value.}
    \label{fig1}
\end{figure*}

\section*{Results}
\subsection*{The HCT outperformed AI advice}

We first compared the performance of the HCT against the performance of the AI-as-advisor approach across the 10 datasets (see Table \ref{datasets} and Methods for details). 
Four datasets are from the domain of medical diagnoses: Two involve diagnoses of cancerous skin lesions by dermatologists \citep{vodrahalli2022humans, chanda2024dermatologist} and two involve detecting lesions during colonoscopy procedures, one diagnosed by experts (at least 5 years of experience in performing colonoscopies) and one by non-experts \citep{reverberi2022experimental}. 
Three datasets consist of judgments of online reviews: One on positive versus negative sentiment classification in beer reviews, one on sentiment classification in book reviews \citep{bansal2021does}, and one on trustworthiness classification of hotel reviews  \citep{lai2019human}. 
Two datasets are from the domain of misinformation detection; specifically, the detection of deepfakes \citep{groh2022deepfake} and truthfulness judgments of news headlines. Finally, one dataset is from the domain of criminal recidivism, in which participants predicted the likelihoods of rearrests \citep{fogliato2021impact}. 
Across datasets, AI advice was provided as discrete choices, as choices with confidence indication, or as class probability predictions.
Altogether, we analyzed over 41,000 judgments made by 1,229 human decision-makers across 3,220 cases (Table \ref{datasets}). 

For every case in every dataset, we tested the accuracy of the HCT by considering the AI advice and generating all possible pairwise permutations of two human decision-makers (unaided by AI advice). Whenever the first human and the AI agreed, we accepted their consensus answer. If they disagreed, a second human was consulted to break the tie. We then calculated the frequency of correct final choices within every case and averaged the HCT performance for a dataset across all cases. We used a similar procedure for the AI-as-advisor approach: For each case within each dataset we calculated the frequency of correct final choices humans made with AI advice, then averaged across cases. For comparison, we also determined human performance without AI advice using the same method. 

 The HCT outperformed the AI-as-advisor approach in all datasets, with improvements ranging from 0.2 to 6.6 percentage points (Fig. \ref{fig1}B). We tested the credibility of these results using a Bayesian estimation and a region of practical equivalence (ROPE) approach of $\pm$ 1 percentage point \citep{benavoli2017time} (see Methods for details). Pooling across all datasets, the HCT improved over the AI-as-advisor approach by 4.45 percentage points (95\% highest density interval [HDI] 3.73--5.27), with 100$\%$ chance of being practically significant (i.e., the proportion of the posterior larger than the ROPE). Table S1 presents dataset-specific results, showing greater than chance practical significance of accuracy improvements in all datasets except for the beer reviews. Fig. S1 presents the same analyses broken down into true positive and true negative rates, showing that the HCT performed better than the AI-as-advisor approach in terms of true positive rate in all datasets, and in half of the datasets it also performed better in terms of true negative rate.

Fig. S2 shows how both methods compare against AI performance directly. When the AI-as-advisor approach performed worse than AI alone, the HCT decreased the performance gap to the AI; in situations where the AI-as-advisor approach performed better than AI alone, the HCT further boosted human--AI accuracy.

These improvements come at a cost: Whenever the first human and AI disagree in the HCT, a second human tiebreaker is needed. We calculated the tiebreaking rate per case over all possible human pairs and averaged across cases within each dataset. The HCT triggered tiebreaking rates in 22$\%$ (book reviews) to 49$\%$ (deception detection) of cases (Table S2). We return to this cost increase in the Discussion section.

\subsection*{The HCT outperformed AI advice with explanation in most realistic scenarios}

A common approach to improving the decision-making accuracy of the AI-as-advisor approach is to make the AI system's decision-making process transparent to human decision-makers. Four of the 10 datasets included one or more condition in which the AI system also provided an explanation of its judgment \citep{bansal2021does, lai2019human, chanda2024dermatologist} (see Methods for details). Combined, these datasets tested 16 explainable AI conditions, containing an additional 50,390 decisions made by 1,423 humans on 516 cases (see Table \ref{datasets_xai} for details per dataset).

Fig. \ref{fig1}C shows the performance of the HCT and the XAI-as-advisor approach across the 16 conditions. Testing for the credibility of these differences with the 1$\%$ ROPE approach showed that the HCT was more accurate in 11 of 16 comparisons with the XAI-as-advisor approach (Table S3). Of special interest is the dataset with expert dermatologists \citep{chanda2024dermatologist}: Here, the XAI-as-advisor approach did not improve over the AI-as-advisor approach but the HCT allowed decision-makers to realize an increase in accuracy of 3 percentage points. In a further two comparisons the HCT and the XAI-as-advisor performed similarly, and in three instances the HCT performed worse. These instances of the HCT performing worse came from the study on deception detection in hotel reviews \citep{lai2019human}: Here, humans on their own were only slightly better than chance (51.1$\%$ accuracy) at detecting deceptive reviews. The strong negative impact this had on the tiebreaking performance of the HCT constrained its performance.

\subsection*{The HCT performed better for all levels of expertise}

\begin{figure*}[!t]
    \centering
    \includegraphics[width = \textwidth]{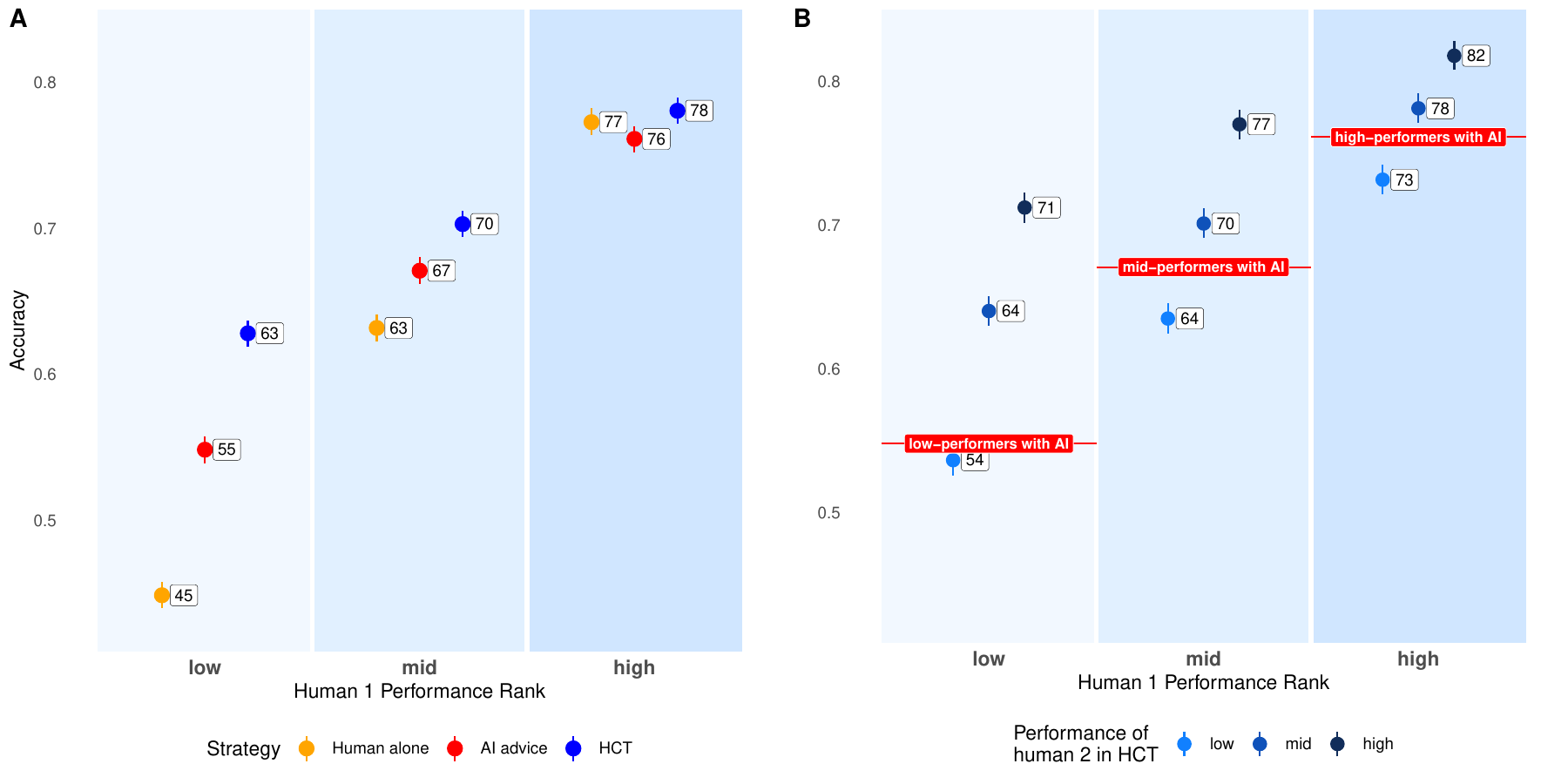}
    \caption{Performance comparison of the hybrid confirmation tree (HCT), the AI-as-advisor approach, and humans alone, as a function of human accuracy. (A) For the HCT, the results are averages given the first individual's performance level; tiebreakers could be low, mid, or high performers. (B) Accuracy of the HCT (dots) and the AI-as-advisor approach (red line) for different levels of human accuracy and different performance levels of the tiebreaker in the HCT. Results are model estimates across the five datasets that used a within-participant design. Point estimates are based on our model; error bars correspond to the 95$\%$ HDI. Numbers show accuracy values.}
    \label{fig2}
\end{figure*}

A central question in human--AI collaboration concerns how AI affects the performance of people with varying levels of expertise \citep{yu2024heterogeneity}. Although poorly performing or less-experienced decision-makers may have a harder time distinguishing between correct and incorrect advice \citep{dratsch2023automation}, they stand to gain more from AI advice than do high-performing decision-makers \citep{tschandl_humancomputer_2020, gaube2023non}---which can potentially close performance gaps between low and high performers \citep{brynjolfsson2025generative}. Predicting the impact of the AI-as-advisor approach and the HCT on individuals with different expertise levels may therefore be challenging. 

To investigate how individual-level performance affects the gains from the HCT and the AI-as-advisor approach, we classified human decision-makers, within each dataset, into three equally sized categories of high, mid, and low performance based on their accuracy without AI advice. We restricted this analysis to the five studies that used a within-subject design in which the same individuals made choices on the same stimuli with and without AI advice: detecting skin cancer \citep{chanda2024dermatologist, vodrahalli_humans_2022}, detecting lesions in colonoscopy procedures \citep{reverberi_experimental_2022}, detecting deepfakes in videos \citep{groh_deepfake_2022}, and predicting rearrest \citep{fogliato_impact_2021}. We paired each individual (from the high-, mid-, or low-performance group) with all other individuals in the HCT. Using Bayesian estimation, we compared how well a decision-maker of a given skill level performed alone, with AI advice, and in the HCT when paired with a low, mid, or high performer.

Better performance in a task without AI support was highly predictive of people's accuracy with an AI advisor as well as within the HCT (Fig. \ref{fig2}A). Importantly, across all levels of human performance, the AI-as-advisor approach outperformed humans alone but the HCT was superior, beating the AI-as-advisor approach by 8 percentage points for low performers (HDI 6.9--9.2), 3.2 percentage points for mid performers (HDI 1.9--4.6), and 1.9 percentage points for high performers (HDI 0.6--3.2; 92$\%$ practically significant). These results also largely held within datasets (Table S4). 

Fig. \ref{fig2}B shows the performance of the HCT as a function of the performance of the first and second individuals in the tree. Unsurprisingly, the performance of the HCT increased with increasing accuracy of the first and second individuals. Independent of their own performance, individuals benefited more from the HCT than from receiving AI advice when paired with a mid or high performer, but were worse off in the HCT when paired with a low performer. However, the percentage point gains in accuracy of being paired with a mid performer (for low performer: 9.3, HDI 8.3--10; mid performer: 3.3, HDI 2.3--4.3; high performer: 2.1, HDI 1--3.1) or a high performer (for low performer: 17, HDI 16--18.1; mid performer: 9.7, HDI 8.7--10.1; high performer: 5.6, HDI 4.6--6.7) were substantially larger than the losses when the first individual was paired with a low performer (for low performer: 1, HDI 0--2; mid performer: 3.3, HDI 2.3--4.3; high performer: 2.9, HDI 1.9--4). 

As only the minority of cases require a tiebreaker, the second position in the HCT is well suited to more in-demand and expensive high-performing decision-makers. For example, senior employees, such as seasoned diagnosticians in medical units, can take up this role, potentially leading to substantial improvements in the overall predictive performance of their junior colleagues.


\subsection*{Accuracy improvements due to greater agreement with correct AI decisions}

\begin{figure*}[!t]
    \centering
    \includegraphics[width = \textwidth]{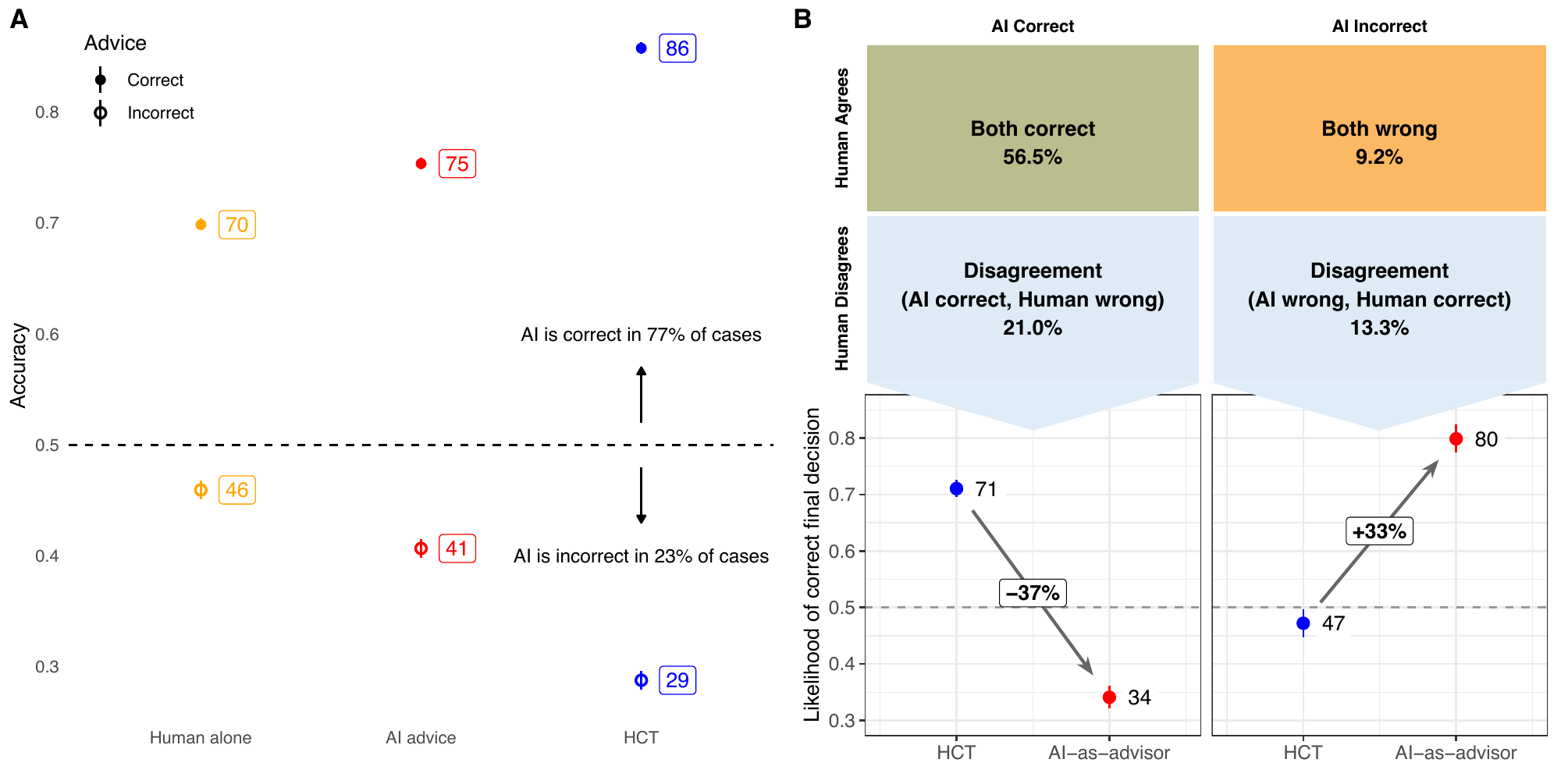}
    \caption{Performance comparison of the hybrid confirmation tree (HCT) and the AI-as-advisor approach, as a function of correct and incorrect AI decisions. (A) Accuracy of the HCT, the AI-as-advisor approach, and humans alone, for cases where the AI was correct and incorrect. (B) Human--AI agreement matrix for correct and incorrect choices (top), showing the performance of the HCT and the AI-as-advisor approach when there was human--AI disagreement and the AI was correct (bottom left) or incorrect (bottom right). Results are model estimates across all datasets; error bars correspond to the 95$\%$ HDI. Numbers show accuracy values.}
    \label{fig3}
\end{figure*}

Research in human factors \citep{parasuraman1997humans, meyer2001effects, yin2025designing, rieger_why_2025} and human--computer interaction \citep{buccinca2021trust, buccinca2025contrastive, vasconcelos2023explanations,guo2024decision} has shown that the main cause of the relatively poor performance of the AI-as-advisor approach is that people cannot distinguish between correct and incorrect AI advice. The effectiveness of the AI-as-advisor approach depends on whether people can correctly decide how to resolve conflict between their own judgment and AI advice---adopting a correct judgment from AI when their own judgment is incorrect (avoiding underreliance), and retaining their own correct judgment when the AI advice is incorrect (avoiding overreliance). The HCT does not require humans to be good at responding to AI advice; instead, it requires humans to be good at solving the decision problem on their own.

To examine why the HCT fared better than the AI-as-advisor approach, we compared how well both methods performed with either correct or incorrect AI advice. We modeled the accuracy of HCT, humans with AI advice and humans without AI advice as a function of whether the AI was correct or incorrect across all 10 datasets. The model was estimated using Bayesian estimation (see Methods for details). 

Fig. \ref{fig3}A shows that when AI was correct the HCT was more frequently correct compared to the AI-as-advisor approach (10.4 percentage points, HDI 9.68--11.15) and to humans without AI advice (15.91 percentage points, HDI 15.16--16.66). However, when AI was incorrect, the performance of the HCT suffered more than the performance of both the AI-as-advisor approach ($-$11.89 percentage points, HDI $-$10.72-- $-$13.15) and humans without AI advice ($-$17.19 percentage points, HDI $-$16.05-- $-$18.42). Given that AI was correct 77\% of the time across all datasets, the benefits from the correct AI advice outweigh the costs of the incorrect AI advice. These results were stable across all datasets (Fig. S3).

To understand the mechanism behind the sensitivity to AI correctness–––where the HCT captures more correct AI decisions but fails to filter out more incorrect ones---we investigated how the HCT and humans with AI advice arrived at their final choice in cases of disagreement (Fig. \ref{fig3}B). In the HCT, disagreement triggers the second human to break the tie. In the AI-as-advisor approach, disagreement occurs when the human's initial choice opposes the AI advice. For the HCT, we calculated the frequency with which the second human tiebreaker selected a correct choice, separately for when AI was correct and when it was incorrect. For the AI-as-advisor approach, we calculated how frequently humans adopted the AI advice that was at odds with their judgment, separately for correct and incorrect AI advice (see Methods for details). 

We once again modeled results using Bayesian estimation, finding that the performance gap stemmed from how each strategy resolves disagreement (Fig. \ref{fig3}). Disagreement arose in about 33\% of cases averaged across our 10 datasets. In the AI-as-advisor approach, humans showed a strong tendency to stick to their initial judgment, regardless of whether the AI judgment was correct or not. Consequently, when the AI judgment was in disagreement and correct (21\% of all cases), humans only adopted it in 34\% of cases. This resistance was beneficial for cases in which AI incorrectly disagreed (13\% of all cases): Humans generally ignored the erroneous advice and stuck to their initial correct choice in 80\% of cases.

The tiebreaker in the HCT resolved conflict differently. When the AI judgment was correct, the tiebreaker agreed with it in 71\% of the cases---twice as frequently as the 34\% uptake of accurate AI advice found in the AI-as-advisor approach. However, independent human--AI choices were similar to each other when AI was incorrect. Here, the tiebreaker rejected incorrect AI advice in only 47\% of cases---much less frequently than the 80\% rejection rate in the AI-as-advisor approach. These tendencies were present in all datasets (Fig. S4). 

In sum, the HCT’s greater accuracy stemmed from two related facts: First, correct AI choices are far more common than incorrect ones, and second, independent human tiebreakers tend to confirm correct AI decisions whereas people are unlikely to change their mind in the face of opposing---but correct---AI advice. And although humans are better at rejecting incorrect AI advice in the AI-as-advisor approach than in the HCT, opposing and incorrect AI advice was not common. As a result, the HCT gained more from confirming correct AI choices than it lost from occasionally endorsing incorrect AI choices.

\subsection*{A signal detection model of AI advice-taking}

Our empirical results demonstrate the superior performance of the HCT, but do not address the cognitive mechanisms underlying the inferior performance of the AI-as-advisor approach. We therefore developed a simple analytical model that tests the theoretical limits of both strategies and can distinguish between two mechanisms that might underlie the inferior performance of the AI-as-advisor approach: i) decision-makers not using enough AI advice, and ii) decision-makers not being able to discriminate between correct and incorrect advice. This model also allowed us to trace different reasons for why the AI-as-advisor approach failed between the datasets.

For simplicity, we assumed for the HCT that decisions among humans and AI are independent and that human decision-makers have the same average performance (see also \citep{berger2026hybridconfirmationtreerobust} for a formal description of HCT performance when human and machine choices are dependent). Let $p_H$ be the human accuracy $p_{AI}$ the accuracy of the AI. The accuracy of the HCT is given by

\begin{align*}
\pi_{\mathrm{HCT}} &= \underbrace{p_H p_{AI}}_{\text{both correct}}
  \notag\\[6pt]
&\quad+\,\underbrace{p_H(1-p_{AI})p_H}_{\substack{\text{first human correct, AI wrong,}\\
                                           \text{second human (tie-breaker) correct}}}
  \notag\\[6pt]
&\quad+\,\underbrace{(1-p_H)p_{AI} p_H}_{\substack{\text{first human wrong, AI correct,}\\
                                           \text{second human correct}}}\,.
\end{align*}

For the AI-as-advisor approach, we introduced two additional parameters, using a signal detection theory (SDT) approach \citep{hautus2021detection}: We followed Langer et al.'s \citep{langer2024effective} conceptualization of AI advice-taking as a signal detection exercise in differentiating correct from incorrect AI advice. We labeled the rate with which humans follow correct AI advice the hit rate (HR) and the rate with which they follow incorrect AI advice as the false alarm rate (FAR). We assumed that humans do not change their mind when their own choice agrees with the AI choice \citep{guo2024decision}. The accuracy of the AI-as-advisor approach is thus given by 

\begin{align*}
\pi_{HAI} &= \underbrace{p_H p_{AI}}_{\text{both correct}} \\
&\quad + \underbrace{p_H(1-p_{AI})(1-\text{FAR})}_{\substack{\text{human correct, AI wrong,} \\ \text{human rejects advice}}} \\
&\quad + \underbrace{(1-p_H)p_{AI} \cdot \text{HR}}_{\substack{\text{human wrong, AI correct,} \\ \text{human takes advice}}} .
\end{align*}

Having calculated hit and false alarm rates, we used the established formulas for SDT---assuming equal variance---to calculate the corresponding discrimination ability (\textit{d'}) and the response bias (\textit{c}; see Methods for details). In our application of SDT, the discrimination ability is the advice-taker's ability to differentiate correct from incorrect AI advice, with higher values indicating a greater ability. The response bias represents an advice-taker's general tendency to accept or reject AI advice, regardless of its quality. A negative bias (\textit{c} $<$ 0) indicates a liberal response, that is, the advice-taker tends to accept AI advice; a positive bias (\textit{c} $>$ 0) indicates a conservative response---the advice-taker tends to reject AI advice. A bias of zero (\textit{c} = 0) means the advice-taker is equally likely to accept or reject AI advice. Fig. S5 presents a visual depiction of this conceptualization. 

The performance of the AI-as-advisor approach changes as a function of the discrimination ability and the response bias (Fig. S6): Higher discrimination ability leads to higher accuracy, and the response bias moderates performance. When humans and AI are similarly accurate, a neutral response bias is best. When AI is more accurate than humans, taking more advice (i.e., a negative response bias) improves performance, and can hurt performance when the human is more accurate than AI.

\begin{figure*}[!t]
    \centering
    \includegraphics[width = \textwidth]{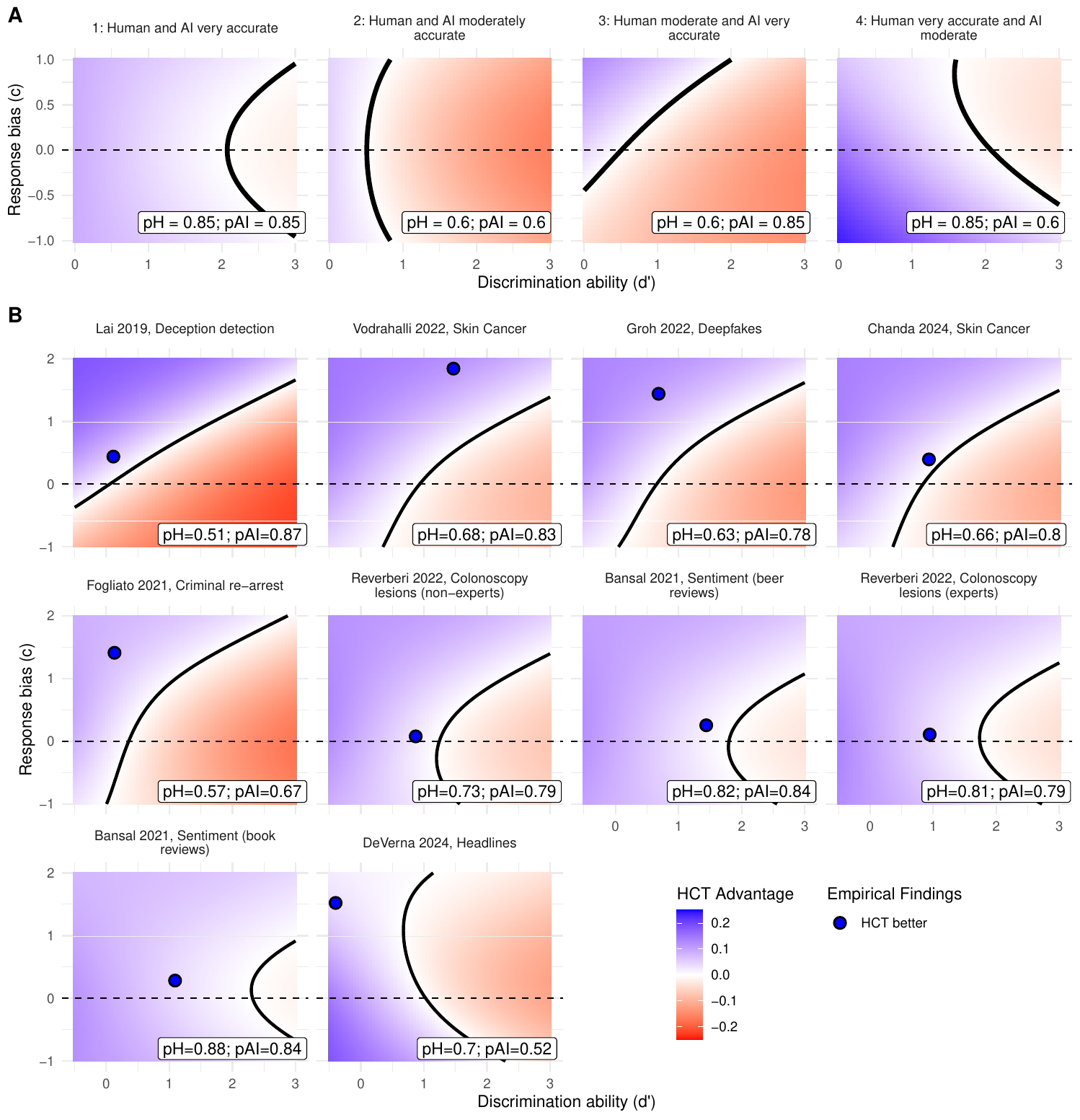}
    \caption{A signal detection model comparing the performance of the hybrid confirmation tree (HCT) and the AI-as-advisor approach. (A) Accuracy difference between the HCT and the AI-as-advisor approach. Solid lines indicate equal accuracy. The accuracy difference is shown as a function of the human ability to discriminate between correct and incorrect AI advice (\textit{d'}) and the overall human propensity to rely on AI advice, expressed as the response bias (\textit{c}). A higher response bias indicates a higher likelihood of not taking advice. Graphs show different combinations of human ($p_H$) and AI accuracy ($p_{AI}$). All values are derived analytically. (B) Empirical results per dataset. Dot color indicates the empirical finding of whether the HCT performed better or worse than the AI-as-advisor approach (Fig. \ref{fig1}B). Panels are ordered by decreasing empirical accuracy difference between the AI alone and humans alone.}
    \label{fig4}
\end{figure*}

Fig. \ref{fig4}A compares the performance of the HCT and the AI-as-advisor approach for different values of discrimination ability and response bias. The specific patterns depend on the underlying human and AI accuracies. When humans and AI had equal individual accuracy, a neutral response bias made high discrimination ability less necessary for achieving equal accuracy between the HCT and the AI-as-advisor approach. Taking a lot or very little advice required a higher discrimination ability for the strategies to perform equally. When both humans and AI were highly accurate, high discrimination ability was required for the AI-as-advisor approach to match the HCT due to the tiebreaker's high performance in the HCT when $p_H$ was high. Conversely, when baseline accuracies of humans and the AI were moderate, even a modest discrimination ability allowed the AI-as-advisor approach to outperform the HCT---the tiebreaker's limited accuracy set a lower bar for the AI-as-advisor approach to be as good as or better than the HCT.
When AI accuracy exceeded human accuracy, a liberal bias (\textit{c} $<$ 0) was increasingly advantageous as discrimination ability improved. Remarkably, even with poor discrimination (low \textit{d'}), liberal acceptance of AI advice could outperform the HCT because the benefit of following the more accurate AI outweighed the cost of following occasional errors. The HCT was constrained by its less accurate human tiebreaker. 
The situation reversed when human accuracy exceeded AI accuracy: The AI-as-advisor approach could only outperform the HCT when the human decision-maker had a high discrimination ability and a conservative bias. Because in this case the HCT could rely on a high $p_H$, tiebreaker accuracy was high. For the AI-as-advisor approach to compete, the human decision-maker had to both maintain a general skepticism towards the AI advice (high \textit{c}) because of its overall low quality and be able to reliably identify the rare instances in which the AI was right and they were wrong (high \textit{d'}).

\subsection*{Empirical validation of our signal detection model}

We applied our signal detection model to our empirical data in order to examine why the HCT was more accurate across the datasets. We first mapped out the analytical expectation of the HCT's advantage or disadvantage based on the empirical accuracies of the human and AI within each dataset, then estimated the empirical discrimination ability and response bias in each dataset (see Methods for details). We found a strong match between our analytical predictions (Fig. \ref{fig4}A) and our empirical results (Fig. \ref{fig1}B): All datasets were in the range of the parameter space where the HCT was predicted to outperform the AI-as-advisor approach. Furthermore, there were stark differences in discrimination ability and response bias across datasets.

Our results provide insights into the unused potential of the AI-as-advisor approach. For instance, in the deception detection dataset \citep{lai2019human}, AI was highly accurate ($p_{AI}$ = 87\%) whereas human accuracy was only slightly better than a coin flip ($p_H$ = 51$\%$). This dataset resembles a situation encountered in our analytical results: Humans are not only held back by low discrimination ability but also by their lack of reliance on AI advice (Fig. \ref{fig4}A). Simply taking more AI advice (i.e., lowering the response bias), which is much more accurate than humans on their own, would have allowed the AI-as-advisor approach to perform as well as or better than the HCT. 

Another situation already encountered in the analytical results is the book sentiment classification dataset, \citep{bansal_does_2021} were both humans and AI exhibited high accuracy ($p_{H}$ = 88\% and $p_{AI}$ = 84\%). In fact, accuracy was so high here that the AI-as-advisor approach could have matched or outperformed the HCT only with a large increase in discrimination ability. Changes to the response bias, which was close to neutral, would have necessitated even larger increases in discrimination ability.

Fig. S7 shows the results of the same analysis using the XAI-as-advisor as comparison (Fig. \ref{fig1}C). Though one might expect an overall increase in discrimination ability and a lower response bias for the XAI-as-advisor as compared to regular AI, there were no clear overall patterns across the datasets. 

Fig. S8 shows the results of the same analysis for high, mid, and low performers (see Fig. \ref{fig3}A). Though one might predict that low performers would benefit more from lower response bias and taking more advice, there were no systematic differences between high and low performers. This result suggests that AI advice is not used adaptively when presented directly. 

In 37 out of 41 comparisons, our model correctly predicted whether the HCT outperforms the AI-as-advisor approach or vice versa. Two errors were made on a practically insignificant differences between HCTs and the XAI-as-advisor approach (Table S3; Lai 2019, Deception detection [Predicted label + random heatmap]; and Bansal 2021, Sentiment [beer reviews; Adaptive {Expert}]). The other two errors were made on differences with weak evidence of practical significance (Lai 2019, Deception detection [Predicted label + examples], PS = 55.4\% in Table S3; and mid performers in Chanda 2024, Skin cancer, PS = 54.7\%).

Our signal detection model is a powerful framework for understanding the accuracy gap between the HCT and the AI-as-advisor approach. It divides human reliance on AI advice into two distinct cognitive mechanisms: the ability to discriminate between correct and incorrect advice and the response bias towards accepting or rejecting it. Our analyses demonstrate that this framework not only explains which strategy will be superior based on the relative accuracies of humans and AI, but also largely explains our empirical findings.

\section*{Discussion}


The HCT is a compelling complementary strategy to the established AI-as-advisor approach for combining human and AI opinions. Across 10 datasets from the domains of medical diagnoses, misinformation detection, sentiment classification, deception detection, and rearrest prediction, the HCT improves decision accuracy compared to the AI-as-advisor approach---even if the AI advice is explained. Furthermore, this benefit does not depend on human expertise: Both low and high performers benefit more from being part of the HCT than from receiving AI advice. We supplemented our analysis with a signal detection model that explains why the HCT is more accurate than the AI-as-advisor approach: Humans cannot discriminate well enough between accurate and inaccurate AI advice and are conservative advice-takers--that is, they rarely change their initial judgments when they disagree with AI systems, even though the AI is on average more accurate than humans alone. The HCT can therefore capitalize more frequently than the AI-as-advisor approach on cases where AI systems make correct choices. This benefit outweighs the losses incurred by the HCT in the rarer cases when AI is incorrect.

A clear downside of the HCT compared to the AI-as-advisor approach is the need to involve a second human to act as a tiebreaker when the human and the AI system disagree. This happens in 20--49\% of cases across our datasets (Table S2). These additional costs need to be weighed against the gains obtained from higher accuracy. 
Consider mammography screening: Do the costs of occasionally consulting an additional radiologist outweigh the reduced need for experts---not to mention the lower emotional toll---that follows from catching more cancers early on and reducing false positives (thereby reducing costly and unnecessary follow-ups)? 

In the HCT, individuals can serve as either the main decision-maker or the tiebreaker according to their expertise. Typically, only a few high performers consistently outperform the average, and engaging them is often costly. Using them as tiebreakers can solve this issue: Expensive experts with little time are only involved in the more challenging cases in which a less experienced human and the AI system disagree. This approach can substantially reduce both overall costs and the strain on in-demand experts (e.g., senior doctors) while improving the performance of low and mid performing individuals (e.g., medical trainees; see Fig. \ref{fig2}). Although different decision problems may require different calculations, our results suggest that there is a large research potential in evaluating the cost of human--AI hybrid systems for organizations \citep{choudhary2025human} or health care systems \citep{eisemann2025nationwide} to make informed choices between AI approaches.

It is still largely unknown when and which human--AI combinations manage to achieve complementarity---that is, when they outperform both humans and AI \citep{vaccaro2024combinations}. The literature on human--AI collaboration predominantly centers on the AI-as-advisor approach; here, humans receive AI suggestions while making their decisions; the independence of judgments is thereby compromised. At the same time, a growing body of work shows the promise of producing complementary human--AI performance by combining independent choices \citep{steyvers2022bayesian, mozannar2023should, zoller2025human}. A common theme amongst these studies is that an algorithm learns to weight and adaptively use more of the human or AI input as deemed necessary given a new decision problem. Research on the wisdom of crowds has used similar algorithms to build weighted ensembles based on the past performance of decision-makers or their decision similarity \citep{kurvers2019detect, budescu2015identifying, mannes2014wisdom,satopaa2014combining, satopaa2017partial, satopaa2023decomposing}. Our research on the HCT contributes to this growing literature on hybrid crowd wisdom by adding a robust strategy that does not require past data for training and parameter tuning, but rather builds on the principle of humans and AI making independent choices \citep{berger2026hybridconfirmationtreerobust}. In that sense, HCTs are similar to fast-and-frugal decision strategies, which require little to no data to reach satisfactory performance levels and avoid the risk of overfitting when a large amount of data is available \citep{gigerenzer1996reasoning}. Research on the majority vote in human crowds---the HCT's closest human-only relative---shows that it is an effective and robust strategy for combining choices. Although it is not the best option all of the time, it is very good most of the time \citep{hastie2005robust}.

A crucial feature of the HCT is that it maintains the independence of human and AI judgments. Findings of automation bias are well-documented across many fields \citep{lyell2017automation, goddard2012automation, manzey2012human, dratsch2023automation} and there is ample evidence that access to digital technology changes how humans learn \citep{loh2016has}. For example, reliance on GPS leads to slower and more flawed acquisition of spatial knowledge \citep{brugger2019does, ishikawa2008wayfinding}. The increasing use of AI in numerous professions---and the widespread integration of generative AI into everyday consumer products---makes it easier for people to offload complex cognitive tasks. While the long-term effects of this phenomenon remain unclear, there is a growing concern about deskilling and learning in the age of AI \citep{natali2025ai, choudhury2024large, lee2025impact}. Early evidence from education \citep{bastani2024generative, darvishi2024impact, wecks2024generative} points towards negative effects of AI use on learning as people start using AI as a crutch; other studies find positive effects of intermittent use of AI \citep{brynjolfsson2025generative} and when humans first attempt to find a solution by themselves \citep{kumar2023math}. Access to large language models (LLMs) may boost individual speed in idea generation; however, the diversity of ideas between humans diminishes compared to human-only generated ideas \citep{doshi2024generative, meincke2025chatgpt, kumar2023math}, confirming concerns that LLMs may impact collective problem solving by correlating individuals' knowledge \citep{burton2024large}. These results emphasize the importance of keeping a functional separation between human and algorithmic reasoning; the HCT may be one way to safeguard human cognition from too much AI influence.

The HCT is a powerful and robust method to successfully combine human and AI choices across a wide variety of real-world high-stakes decisions. Easy to implement, it maintains human agency and allows decision-makers to tap into the potential of hybrid intelligence more consistently when compared to receiving AI-advice.

\section*{Methods}

\subsection*{Data and code availability}
All data and code to reproduce the analyses are available at \url{https://osf.io/dvgh4/?view_only=aa04f5bcf0ed410ab54fdc827c424eec}.

\subsection*{Datasets}

Tables \ref{datasets} and \ref{datasets_xai} summarize the key characteristics of the datasets used in this study. Here we briefly describe the data collection procedure for each dataset.

\begin{table}[ht]
\caption{Overview of datasets without explainable AI: number of binary choices, distinct human raters, and cases per study.}
\centering
\begin{tabular}{@{} l r r r @{}}
\toprule
Dataset & Choices & Humans & Cases \\
\midrule
Bansal 2021, Sentiment (beer reviews)             & 4,650  &  92  &  50  \\
Bansal 2021, Sentiment (book reviews)             & 4,550  &  91  &  50  \\
Chanda 2024, Skin cancer                          & 1,508  & 109  & 196  \\
DeVerna 2024, Headlines                           & 9,640  & 241  &  40  \\
Fogliato 2021, Criminal rearrest                 & 5,753  & 271  & 1,983  \\
Groh 2022, Deepfakes                              & 3,044  & 304  &  54  \\
Lai 2019, Deception detection                     & 1,600  &  80  & 319  \\
Reverberi 2022, Colonoscopy (experts)     & 5,040  &  10  & 504  \\
Reverberi 2022, Colonoscopy (non‑experts) & 5,544  &  11  & 504  \\
Vodrahalli 2022, Skin cancer                      &  480  &  20  &  24  \\
\midrule
\textbf{Total}                                     & \textbf{41,809} & \textbf{1,229} & \textbf{3,220} \\
\bottomrule
\end{tabular}
\label{datasets}
\end{table}

\begin{table}[ht]
\caption{Overview of datasets with XAI-as-advisor conditions Number of binary choices, distinct human raters, and cases per study. Original names of interventions in parentheses.}
\centering
\begin{tabular}{@{}p{0.50\linewidth} r r r @{}}
\toprule
Dataset (condition) & Choices & Humans & Cases \\
\midrule
Bansal 2021, Sentiment (beer reviews) (Adaptive [AI])             &  4,900 &  98 &  50 \\
Bansal 2021, Sentiment (beer reviews) (Adaptive [Expert])         &  5,050 & 100 &  50 \\
Bansal 2021, Sentiment (beer reviews) (Explain-Top-1)             &  4,400 &  88 &  50 \\
Bansal 2021, Sentiment (beer reviews) (Explain-Top-2)             &  4,650 &  92 &  50 \\
Bansal 2021, Sentiment (book reviews) (Adaptive (Expert)         &  4,750 &  95 &  50 \\
Bansal 2021, Sentiment (book reviews) (Adaptive)                  &  4,650 &  93 &  50 \\
Bansal 2021, Sentiment (book reviews) (Explain-Top-1)             &  4,550 &  91 &  50 \\
Bansal 2021, Sentiment (book reviews) (Explain-Top-2)             &  4,500 &  90 &  50 \\
Chanda 2024, Skin cancer (xAI)                                    &  1,740 & 116 & 196 \\
Lai 2019, Deception detection (Examples)                          &  1,600 &  80 & 319 \\
Lai 2019, Deception detection (Heatmap)                           &  1,600 &  80 & 319 \\
Lai 2019, Deception detection (Highlight)                         &  1,600 &  80 & 319 \\
Lai 2019, Deception detection (Predicted label + heatmap)         &  1,600 &  80 & 319 \\
Lai 2019, Deception detection (Predicted label + examples)        &  1,600 &  80 & 319 \\
Lai 2019, Deception detection (Predicted label + random heatmap)  &  1,600 &  80 & 319 \\
Lai 2019, Deception detection (Predicted label w/ accuracy)       &  1,600 &  80 & 319 \\
\midrule
\textbf{Total}                                                    & \textbf{50,390} & \textbf{1,423} & \textbf{516} \\
\bottomrule
\end{tabular}
\label{datasets_xai}
\end{table}

\subsubsection*{Lai 2019; Deception detection}

We used data from \citep{lai2019human}, in which experiments were conducted on Amazon Mechanical Turk. The dataset for deception detection consists of 800 genuine hotel reviews from TripAdvisor and 800 deceptive reviews written by MTurk workers for 20 hotels in Chicago. The authors used 80$\%$ of these reviews to train their machine learning model and the remaining 20$\%$ as their held-out test set for human evaluation. Participants in the study determined whether reviews were genuine or deceptive. Each participant evaluated 20 reviews after a training session and an attention check. We focused only on the individuals who passed the attention check, following the original authors. Various experimental conditions provided different levels of machine assistance, including feature-based explanations (highlighting words or using a heatmap), example-based explanations, machine-predicted labels (with or without an explicit statement of machine accuracy), and combinations thereof. We focused our analysis on the \textit{Control} condition, in which 80 participants labeled the test set (each review received five human labels) without any machine assistance, resulting in 1,600 human predictions. We further analyzed seven treatments in which AI explanations were provided, focusing on conditions where only explanations (heatmap, highlight, examples) were given without predictions, and conditions where predictions were shown in addition to AI explanations and information about AI accuracy. The AI predictions were taken from the original data and are based on a support vector machine. The authors trained a linear support vector machine with bag-of-words features on the 80$\%$ training split of the reviews. This model achieved an accuracy of 87$\%$.

\subsubsection*{Fogliato 2021; Criminal rearrest}

We used data from \citep{fogliato2021impact} based on a vignette experiment. The offender profiles presented to participants were derived from a dataset provided by the Pennsylvania Commission on Sentencing that contained information about offenders sentenced in the state’s criminal courts. From this, a survey sample of 3,523 offender profiles was selected for the study, using stratified random sampling to reflect the test population on race, sex, age, and rearrest status. In the study, 531 laypeople recruited on Amazon Mechanical Turk predicted the likelihood of rearrest as a binary prediction (rearrest or not) for a series of 40 offender profiles. The study employed a between-subjects design to examine anchoring effects: Some participants saw a risk assessment instrument (RAI) prediction upfront (anchoring condition) and others made an initial prediction before seeing the RAI prediction and could revise afterward (non-anchoring condition). We focused our analyses on the non-anchoring condition. Participants saw short descriptions of offenders. Outcome feedback on rearrest was provided for the first 14 offenders, while RAI predictions were shown for the rest. We based our analysis on the latter. Participants made probability predictions as well as binary predictions about rearrest; we used only the binary predictions. We restricted our analyses to offender profiles that were assessed by at least two human participants (a requirement for calculating the HCT), which yielded a set of 1,983 cases. The AI advice used in the study was a Lasso logistic regression model developed by the authors. It was trained on 70$\%$ of the full Pennsylvania sentencing dataset of more than 110,000 profiles to predict 3-year post-release rearrest using demographic features (age, sex, race), current charge information, and prior criminal history. AI's prediction accuracy was 67$\%$ in our selected set of cases.

\subsubsection*{Groh 2022; Deepfakes}

We used the data from \citep{groh2022deepfake}, which consists of 54 videos from the Deepfake Detection Challenge, of which half were deepfakes and half were not \citep{dolhansky2020deepfake}. We focused our analyses on the data collected in Experiment 2: Single Video Design for which participants---recruited on Prolific---indicated whether one video was a deepfake on a slider ranging from ``100\% confidence this is not a deepfake'' to ``100\% confidence this is a deepfake,'' (midpoint: ``just as likely a deepfake as not a deepfake''). Participants' judgments at the midpoint were randomly assigned as either a deepfake or a no deepfake choice. Afterwards, participants saw an AI model's probability predictions and could adjust their prediction. We restricted our analysis to participants in the control conditions who passed the attention check. For the AI classification of the videos, we used the predictions of the winning model in the Deepfake Detection Challenge by \citep{seferbekov2021deepfake} that were provided by the original authors, which were also the predictions provided to participants. This model achieved an accuracy of 78$\%$.

\subsubsection*{Vodrahalli 2022; Skin cancer}

We used data from expert dermatologists presented in \citep{vodrahalli2022humans}. The task involved determining whether a skin lesion, shown as a single image, should be biopsied (i.e., malignant or benign). The images were selected by a board-certified dermatologist from the International Skin Imaging Collaboration (ISIC) database. Participants made an initial assessment and then a revised assessment after receiving advice binary diagnostic advice. A total of 37 dermatologists participated, recruited through public outreach. The experiment followed a between-subjects design: Dermatologists were randomly assigned to receive advice attributed to either AI or a peer group. We used only the condition in which people received AI advice.  Dermatologists provided their responses on a continuous sliding scale indicating confidence in benign versus malignant. We binarized their choices, randomly assigning midpoint choices to either benign or malignant. This advice was generated using a ResNet-18 model trained by the authors on the ISIC dataset to predict malignancy. Across the data sample shown in the task, the AI model had an accuracy of 83$\%$.

\subsubsection*{Chanda 2024; Skin cancer}

We used data from a study by \citep{chanda2024dermatologist} that developed and evaluated an explainable AI system for diagnosing melanoma from dermoscopic images. The authors focused on biopsy-verified melanoma and nevus images, resulting in a base set of 3,611 images from 1,981 unique lesions. To train their AI model, the authors first acquired ground-truth annotations for these 3,611 images from 14 international board-certified dermatologists. The annotators, who were aware of the ground truth diagnosis (melanoma or nevus), explained the diagnosis by selecting relevant features from a compiled dermascopic ontology and annotating the corresponding regions of interest on the images. The base set was then split into training (2,646 images, 1,460 lesions), validation (599 images, 321 lesions), and a test set (200 unique lesions: 100 melanomas, 100 nevi, only one image per lesion). The explainable AI model, using a ResNet50 backbone, was trained on these expert annotations to predict lesion characteristics and infer a diagnosis. The model achieved a balanced accuracy of 80$\%$ on the test set. For the human evaluation, a three-phase reader study was conducted with 116 international clinicians (dermatologists). Phase 1 (No AI): Clinicians diagnosed 15 lesions (14 unique from the test set, one repeated), selected explanations from the ontology, annotated regions of interest, and stated their diagnostic confidence. Phase 2 (AI Support): Clinicians diagnosed the same lesions again, this time with access to the AI's diagnosis (melanoma/nevus) and its sensitivity/specificity, but without explanations. They provided their diagnosis, confidence, and trust in the AI. Phase 3 (XAI Support): Clinicians diagnosed the lesions a final time, now with full explainable AI support---the AI's diagnosis, its sensitivity/specificity, and textual and localized visual explanations for the AI's detected characteristics (based on Grad-CAM saliency maps and the ontology). They again provided their diagnosis, confidence, and trust.

\subsubsection*{DeVerna 2024; Headlines}

We used data from an experiment by \citep{deverna2024fact} investigating the impact of LLM-generated fact-checking information on belief in, and intent to share, political news headlines. The stimuli consisted of 40 political news stories, each with a headline, a lead sentence (if present), and an image. Half of the headlines were true and half were false; within each true or false category, half were favorable to Democrats and half to Republicans. The study recruited a representative sample of 2,159 U.S. participants via Qualtrics. Participants were randomly assigned to either a belief group (asked if they believed headlines were accurate) or a sharing group (asked if they would share them). We investigated the belief group. Within each group, participants were further assigned to one of several conditions. We focused on the control group (no AI) and the LLM-forced condition, in which participants saw the headlines along with fact-checking information generated by ChatGPT 3.5 and indicated whether they believed the headline was accurate. The LLM's responses were then categorized by the original authors as judging the headline as true, false, or unsure. The LLM accurately identified 90$\%$ of false headlines  but was unsure about 65$\%$ of true headlines and mislabeled 20$\%$ of true headlines. If the model was unsure, we assigned an incorrect classification, following the original authors. This resulted in a model accuracy of 52$\%$.

\subsubsection*{Reverberi 2022; Colonoscopy lesions}

We analyzed data from an experimental study by \citep{reverberi2022experimental} investigating human--AI collaboration in the optical diagnosis of colorectal lesions during colonoscopies. The stimuli consisted of 504 video clips of real colonoscopies, each presenting one lesion.  The histopathological diagnosis of each lesion served as the ground truth. Twenty-one endoscopists participated in the experiment: 10 experts (at least 5 years of colonoscopy experience and experience in optical biopsy with virtual chromoendoscopy) and 11 non-experts (fewer than 500 colonoscopies performed). The study employed a within-subjects design with two sessions. Session 1 (No AI Advice): Endoscopists diagnosed the 504 lesions without diagnostic AI advice. The AI system only highlighted the target lesion with a green box. For each lesion, participants provided a categorical diagnosis (later mapped to \textit{adenoma}, \textit{non-adenoma}, or \textit{uncertain}). Session 2 (AI Advice): At least 2 weeks after Session 1 for each batch of videos, endoscopists diagnosed the same 504 lesions again, this time with the AI system providing diagnostic advice (\textit{adenoma}, \textit{non-adenoma}, and \textit{undefined}) in addition to highlighting the lesion. Participants again provided their diagnosis. We treated all uncertain diagnoses by humans and AI as incorrect diagnoses, following the original authors. This led to an AI accuracy of 79$\%$.

\subsubsection*{Bansal 2021, Beer and book reviews}

We used data from a study conducted by \citep{bansal2021does} that investigated the effect of AI explanations on human decision-making in sentiment analysis tasks. The study used two datasets: beer reviews and Amazon book reviews. For each dataset, 50 unambiguous reviews (as judged by human participants in a pilot study) were selected from a larger test set such that the AI's accuracy on these samples was comparable to human accuracy. Participants were recruited from MTurk. After a screening phase, approximately 100 participants per condition remained. The main task involved participants predicting the sentiment (positive/negative) of the 50 selected reviews. This was a between-subjects design in which participants were assigned to different experimental conditions of AI assistance that was provided concurrently to participants solving the task. Our reanalysis focuses on the following conditions (including original study names): Team (Conf): Participants saw the AI's sentiment prediction (positive/negative) and its confidence score, but no explanation; Team (Explain-Top-1, AI): In addition to the AI's prediction and confidence, participants saw an AI-generated explanation highlighting the most influential words for the predicted sentiment class; Team (Explain-Top-2, AI): Similar to Explain-Top-1, but the AI explanation highlighted influential words for the top two predicted sentiment classes (e.g., positive and negative); Team (Adaptive, AI): The AI provided Explain-Top-1 if its confidence in the classification was above a threshold; Team (Adaptive, Expert): Similar to Team (Adaptive, AI) in logic, but the explanations (highlighted phrases) were generated by a human expert. The AI still provided the initial prediction and confidence. The AI model used for sentiment classification and generating predictions was a RoBERTa-based text classifier fine-tuned on the respective datasets. Participants received immediate feedback on their correctness after each review. The AI accuracy for both datasets was 84$\%$.

\subsection*{Bayesian estimation}

We used Bayesian estimation to compare the difference in accuracy between the HCT and the AI-as-advisor approach across the datasets. We relied on the general procedure described by \citep{benavoli2017time} for building generalized linear mixed models (GLMMs) to account for correlated samples and test for a difference using a ROPE approach of $\pm$ 1 percentage point. We implemented this strategy using \texttt{brms} \citep{burkner2017brms} in R 4.4.2 \citep{rcore} and made extensive use of the \texttt{tidyverse} package \citep{wickham2019welcome}. We ran all models with four chains and 10,000 iterations after warm-up and checked chain convergence using the Gelman--Rubin criterion $\hat{R}$ $<$ 1.01. Using the fitted GLMMs, we applied \texttt{marginaleffects} \citep{arel2024interpret} to compute the contrasts between factors when necessary, and described the median contrast and 95$\%$ HDI between conditional marginal posteriors. We used \texttt{bayestestR} \citep{makowski2019bayestestr} to compute the probability of direction (percentage of the posterior difference that is greater than 0) and the practical significance (percentage of the posterior difference that is greater than the the upper limit of the ROPE).

To test differences in accuracy between the HCT and the AI-as-advisor or XAI-as-advisor approach (Figs. \ref{fig1}, S2), we used a GLMM with no-intercept and hierarchical structure of cases contained within datasets using the formula \texttt{accuracy $\sim$ 0 + strategy + (1 | dataset / case)}, where strategy is a factor indicating either the HCT or the AI-as-advisor or XAI-as-advisor approach and accuracy is the accuracy achieved by either strategy on a case. When we modeled the accuracy of the strategies within a dataset, we omitted the dataset level from the hierarchy specification in the formula interface. 

We used the setup again to model how individuals benefit from either using AI advice or being part of the HCT and being paired with varying levels of human performance (Fig. \ref{fig2}). For Fig. \ref{fig2}A we modeled the absolute accuracy of the strategies as an interaction between strategy and the first human performance level using the formula 
\texttt{accuracy $\sim$ 0 + strategy * performance level of human 1 + (1 | dataset / human 1)}. For Fig. \ref{fig2}B, we modeled the accuracy of the HCT with \texttt{HCT accuracy $\sim$ 0 + performance level of human 1 * performance level of human 2 + (1 | dataset / human 1)}. To compare the accuracy of different combinations of human performance levels in the HCT with the accuracy of the AI-as-advisor approach, we modeled the difference in accuracy as \texttt{difference in accuracy $\sim$ 0 + performance level of human 1 * performance level of human 2 + (1 | dataset / human 1)}. The hierarchical structure accounts for variance between humans nested in datasets.

We also used Bayesian estimation to describe the data in Fig. \ref{fig3}. For Fig. \ref{fig3}A, we fit a linear GLMM \texttt{correct $\sim$ 0 + modelcorrect * strategy + (1| dataset/case)} on all cases of all datasets where \texttt{correct} is the accuracy for a given case and this accuracy is modeled using and interaction between \texttt{modelcorrect} (i.e., whether the AI made a correct or incorrect choice) and \texttt{strategy} (i.e., whether humans alone, humans with AI or the HCT produced the choice). In Fig. \ref{fig3}B, we fit a linear GLMM \texttt{finalchoice $\sim$ 0 + modelcorrect * strategy + (1| dataset/case)}, where the outcome is the likelihood of making a correct final choice on all cases where there was no human--AI agreement.

\subsection*{Calculating AI reliance}

Fig. \ref{fig3}B shows the rate of humans agreeing with AI advice when it was correct and incorrect. This rate is conditional on disagreement between the human's initial decision and the AI recommendation, as suggested by \citep{guo2024decision}.

For studies employing a within-subjects design \citep{chanda2024dermatologist, reverberi2022experimental, groh2022deepfake, vodrahalli2022humans, fogliato2021impact}, we directly observed each participant's response to AI advice. We first identified all instances where the participant's initial decision disagreed with the AI recommendation, then calculated the acceptance rate separately for cases where AI was correct and incorrect. These rates were calculated for each individual participant and then averaged within each dataset to obtain average estimates.

For the studies employing a between-subjects design \citep{bansal_does_2021, lai2019human, deverna2024fact}, we inferred acceptance behavior by comparing the decisions of humans in the no-AI (control) group with decisions from humans receiving AI advice. For each case in a dataset, we first calculated the proportion of control group participants who disagreed with the AI recommendation, representing the potential for change. We then measured the actual change by computing the difference in average decisions between the control group and the AI advice group. When the control group unanimously agreed with the AI (no potential for change), we set the acceptance rate to zero. Similarly, when participants moved away from the AI recommendation (indicated by negative change or ratios exceeding 1), we set the acceptance rate to zero, which indicates rejecting rather than accepting advice. These case-level rates were then averaged within each dataset, separately for correct and incorrect AI recommendations.

\subsection*{Signal detection model}

To model the AI-as-advisor approach as a signal detection problem, we assumed a standard signal detection setup with equal variance between the signal and noise distributions \citep{hautus2021detection} (see also Fig. S4).

We calculated the discrimination ability \textit{d'} as

\[
d' = z(HR) - z(FAR),
\]
where \textit{z} is the z-score, HR is the hit rate, and FAR is the false alarm rate. The response bias \textit{c} is defined as

\[
c = -\tfrac{1}{2}\big[z(HR) + z(FAR)\big].
\]

As indicated in the main text, we calculated hit rates and false alarm rates to characterize how participants responded to AI advice when it was correct versus incorrect. The hit rate represents the probability of accepting correct AI advice; the false alarm rate represents the probability of accepting incorrect AI advice. Both rates are conditional on disagreement between the human's initial decision and the AI recommendation.

For studies employing a within-subjects design \citep{chanda2024dermatologist, reverberi2022experimental, groh2022deepfake, vodrahalli2022humans, fogliato2021impact}, we directly observed each participant's response to AI advice. We first identified all instances where the participant's initial decision disagreed with the AI recommendation. Among these cases, we calculated the acceptance rate separately for cases where the AI was correct (hit rate) and cases where it was incorrect (false alarm rate). These rates were calculated for each participant then averaged within each dataset to obtain average estimates.

For the studies employing a between-subjects design \citep{bansal_does_2021, lai2019human, deverna2024fact}, we inferred acceptance behavior by comparing the decisions of humans in the no-AI (control) group with decisions from humans receiving AI advice. For each case in a dataset, we first calculated the proportion of control group participants who disagreed with the AI recommendation, representing the potential for change. We then measured the actual change by computing the difference in average decisions between the control group and the AI advice group. For cases where the AI was correct, the ratio of actual change to potential change provided the hit rate; for cases where the AI was incorrect, this ratio provided the false alarm rate. When the control group unanimously agreed with the AI (no potential for change), we set the acceptance rate to zero. Similarly, when participants moved away from the AI recommendation (indicated by negative change or ratios exceeding 1), we set the acceptance rate to zero, as this indicates rejecting advice. These case-level rates were then averaged within each dataset, separately for correct and incorrect AI recommendations.

\bibliographystyle{plainnat} 
\bibliography{_references} 

\newpage
\appendix 

\renewcommand{\figurename}{Fig S}
\renewcommand{\tablename}{Table S}

\setcounter{figure}{0}
\setcounter{table}{0}

\begin{table}[ht]
\centering
\caption{Median contrast between the marginal posterior estimates of between the hybrid confirmation tree and humans with AI advice, 95\% HDI, and diagnostic summaries (Probability of direction (PD), Probability of significance (PS), \(\widehat{R}\)) for each data set.}
\resizebox{\textwidth}{!}{%
\begin{tabular}{@{}lrrrrrr@{}}
\toprule
Data set & Contrast & HDI Low & HDI High & PD & PS & $\widehat R$ \\
\midrule     
Chanda 2024, Skin Cancer                          & 0.0334  & 0.0103  & 0.0569  & 0.998 & 0.976 & 1.00 \\
Fogliato 2021, Criminal re-arrest                 & 0.0583  & 0.0461  & 0.0706  & 1.000 & 1.000 & 1.00 \\
DeVerna 2024, Headlines                           & 0.0300  & -0.0417 & 0.0994  & 0.804 & 0.716 & 1.00 \\
Lai 2019, Deception detection                     & 0.0665  & 0.0303  & 0.1030  & 1.000 & 0.999 & 1.00 \\
Reverberi 2022, Colonoscopy lesions (experts)     & 0.0223  & 0.0093  & 0.0348  & 1.000 & 0.972 & 1.00 \\
Reverberi 2022, Colonoscopy lesions (non-experts) & 0.0113  & -0.0012 & 0.0253  & 0.955 & 0.574 & 1.00 \\
Vodrahalli 2022, Skin Cancer                      & 0.0467  & -0.0333 & 0.1210  & 0.888 & 0.832 & 1.00 \\
Groh 2022, Deepfakes                              & 0.0456  & 0.0110  & 0.0776  & 0.996 & 0.981 & 1.00 \\
Bansal 2021, Sentiment (book reviews)             & 0.0219  & 0.0098  & 0.0340  & 1.000 & 0.972 & 1.00 \\
Bansal 2021, Sentiment (beer reviews)             & 0.0027  & -0.0173 & 0.0227  & 0.598 & 0.229 & 1.00 \\
\bottomrule
\end{tabular}%
}
    \label{SI_table1}
\end{table}

\begin{table}[ht]
\centering
\caption{Average number of decision makers (cost) used by the hybrid confirmation tree per data set.}
\begin{tabular}{@{}l r@{}}
\toprule
Data set & \(n\) \\
\midrule
Bansal 2021, Sentiment (beer reviews)               & 1.30 \\
Bansal 2021, Sentiment (book reviews)               & 1.22 \\
Chanda 2024, Skin Cancer                            & 1.33 \\
DeVerna 2024, Headlines                             & 1.48 \\
Fogliato 2021, Criminal re--arrest                   & 1.38 \\
Groh 2022, Deepfakes                                & 1.42 \\
Lai 2019, Deception detection                       & 1.49 \\
Reverberi 2022, Colonoscopy lesions (experts)       & 1.20 \\
Reverberi 2022, Colonoscopy lesions (non--experts)   & 1.25 \\
Vodrahalli 2022, Skin Cancer                        & 1.35 \\
\bottomrule
\end{tabular}%
    \label{table2_cost}
\end{table}

\newpage

\begin{table}[ht]
\centering
\caption{Median contrast between the marginal posterior estimates of between the hybrid confirmation tree and humans with explainable AI advice, 95\% HDI, and diagnostic summaries (Probability of direction (PD), Probability of significance (PS), \(\widehat{R}\)) for each explainable AI treatment.}
\resizebox{\textwidth}{!}{%
\begin{tabular}{@{}lrrrrrr@{}}
\toprule
Data set (treatment) & Contrast & HDI Low & HDI High & PD & PS & $\widehat R$ \\
\midrule
Lai 2019, Deception detection (Highlight) & 0.1248 & 0.0869 & 0.1697 & 1.000 & 1.000 & 1.00 \\
Lai 2019, Deception detection (Heatmap) & 0.1097 & 0.0725 & 0.1488 & 1.000 & 1.000 & 1.00 \\
Lai 2019, Deception detection (Examples) & 0.1402 & 0.1009 & 0.1789 & 1.000 & 1.000 & 1.00 \\
Lai 2019, Deception detection (Predicted label w/ accuracy) & -0.0601 & -0.0926 & -0.0246 & 0.999 & 0.998 & 1.00 \\
Lai 2019, Deception detection (Predicted label + heatmap) & -0.0405 & -0.0733 & -0.0081 & 0.991 & 0.963 & 1.00 \\
Lai 2019, Deception detection (Predicted label + examples) & -0.0119 & -0.0406 & 0.0205 & 0.789 & 0.554 & 1.00 \\
Lai 2019, Deception detection (Predicted label + random heatmap) & -0.0066 & -0.0422 & 0.0287 & 0.644 & 0.420 & 1.00 \\
Chanda 2024, Skin Cancer (xAI) & 0.0259 & 0.0052 & 0.0471 & 0.990 & 0.928 & 1.00 \\
Bansal 2021, Sentiment (beer reviews) (Explain-Top-2) & 0.0215 & -0.0037 & 0.0470 & 0.947 & 0.814 & 1.00 \\
Bansal 2021, Sentiment (beer reviews) (Explain-Top-1) & 0.0118 & -0.0166 & 0.0375 & 0.794 & 0.547 & 1.00 \\
Bansal 2021, Sentiment (beer reviews) (Adaptive (AI)) & 0.0104 & -0.0139 & 0.0338 & 0.807 & 0.514 & 1.00 \\
Bansal 2021, Sentiment (beer reviews) (Adaptive (Expert)) & -0.0005 & -0.0212 & 0.0206 & 0.520 & 0.183 & 1.00 \\
Bansal 2021, Sentiment (book reviews) (Explain-Top-2) & 0.0187 & -0.0012 & 0.0381 & 0.970 & 0.811 & 1.00 \\
Bansal 2021, Sentiment (book reviews) (Explain-Top-1) & 0.0105 & -0.0012 & 0.0224 & 0.959 & 0.525 & 1.00 \\
Bansal 2021, Sentiment (book reviews) (Adaptive) & 0.0113 & -0.0059 & 0.0293 & 0.887 & 0.549 & 1.00 \\
Bansal 2021, Sentiment (book reviews) (Adaptive (Expert)) & 0.0185 & -0.0025 & 0.0390 & 0.959 & 0.803 & 1.00 \\
\bottomrule
\end{tabular}%
}
\label{SI_table_xAI_marginals}
\end{table}

\newpage

\begin{table}[ht]
\centering
\caption{Median contrast between the marginal posterior estimates of between the hybrid confirmation tree and humans with AI advice for every performance rank, 95\% HDI, and diagnostic summaries (Probability of direction (PD), Probability of significance (PS), \(\widehat{R}\)) for each data set.}
\resizebox{\textwidth}{!}{%
\begin{tabular}{@{}llrrrrrr@{}}
\toprule
Data set & Performance Rank & Contrast & HDI Low & HDI High & PD & PS & $\widehat R$ \\
\midrule
Chanda 2024, Skin Cancer & low & -0.0012 & -0.0310 & 0.0291 & 0.531 & 0.282 & 1.00 \\
 & mid & 0.0119 & -0.0187 & 0.0427 & 0.774 & 0.547 & 1.00 \\
 & high & 0.0844 & 0.0547 & 0.1160 & 1.000 & 1.000 & 1.00 \\
\addlinespace
Reverberi 2022, Colonoscopy lesions (non-experts) & low & 0.0163 & -0.0171 & 0.0496 & 0.841 & 0.649 & 1.00 \\
 & mid & 0.0104 & -0.0224 & 0.0442 & 0.736 & 0.510 & 1.00 \\
 & high & 0.0096 & -0.0271 & 0.0502 & 0.696 & 0.491 & 1.00 \\
\addlinespace
Reverberi 2022, Colonoscopy lesions (experts) & low & 0.0412 & 0.0193 & 0.0618 & 1.000 & 0.996 & 1.00 \\
 & mid & 0.0167 & -0.0072 & 0.0413 & 0.914 & 0.709 & 1.00 \\
 & high & 0.0045 & -0.0204 & 0.0295 & 0.644 & 0.324 & 1.00 \\
\addlinespace
Vodrahalli 2022, Skin Cancer & low & 0.0487 & 0.0139 & 0.0839 & 0.996 & 0.984 & 1.00 \\
 & mid & 0.0468 & 0.0118 & 0.0817 & 0.995 & 0.979 & 1.00 \\
 & high & 0.0479 & 0.0093 & 0.0855 & 0.992 & 0.974 & 1.00 \\
\addlinespace
Groh 2022, Deepfakes & low & 0.0842 & 0.0614 & 0.1060 & 1.000 & 1.000 & 1.00 \\
 & mid & 0.0397 & 0.0170 & 0.0624 & 1.000 & 0.994 & 1.00 \\
 & high & 0.0086 & -0.0142 & 0.0315 & 0.771 & 0.452 & 1.00 \\
\addlinespace
Fogliato 2021, Criminal re-arrest & low & 0.1150 & 0.0993 & 0.1320 & 1.000 & 1.000 & 1.00 \\
 & mid & 0.0315 & 0.0149 & 0.0478 & 1.000 & 0.994 & 1.00 \\
 & high & 0.0038 & -0.0124 & 0.0199 & 0.680 & 0.230 & 1.00 \\
\bottomrule
\end{tabular}%
}
\label{SI_table_expertise}
\end{table}

\newpage

\begin{figure*}[!htbp]
    \centering
    \includegraphics[width = \textwidth]{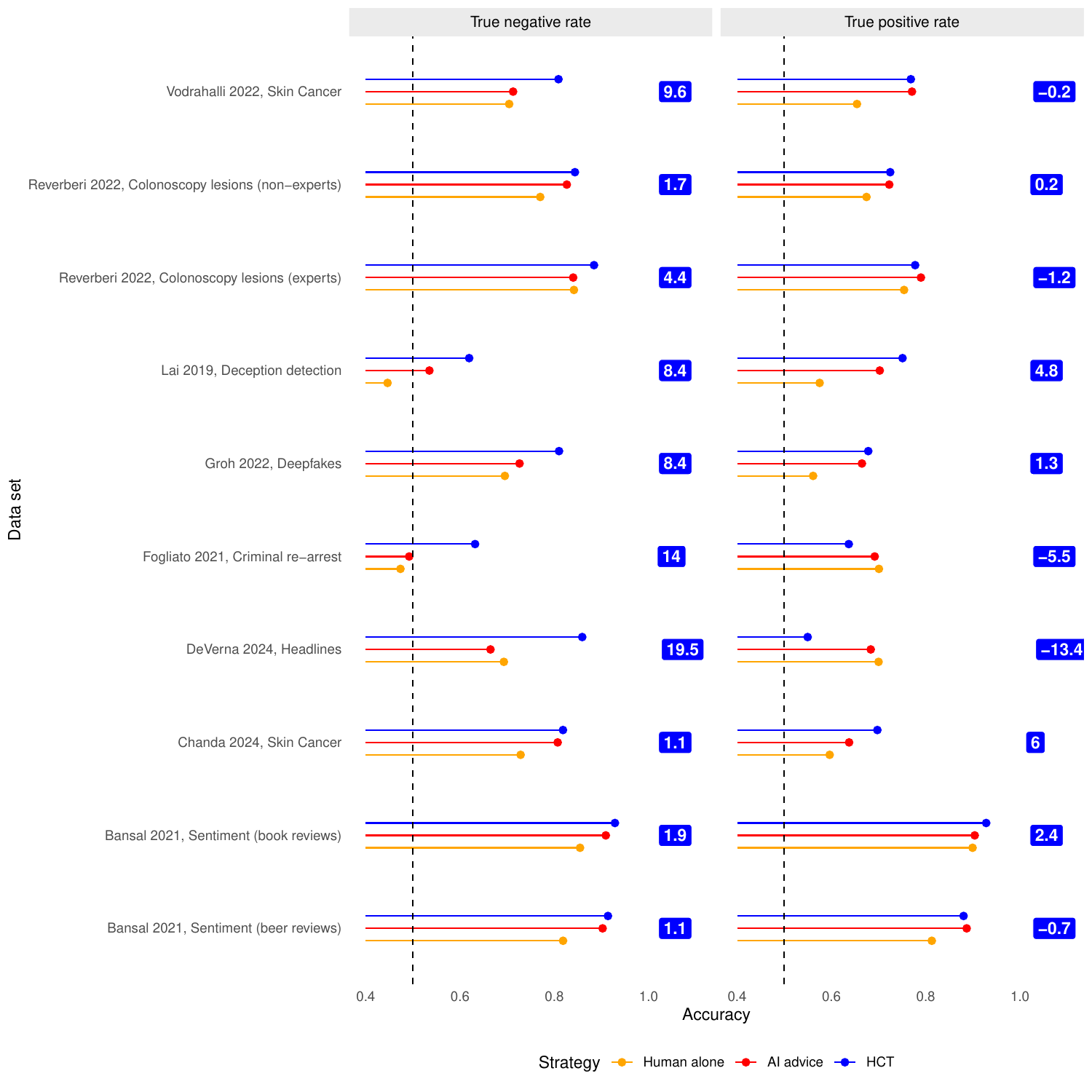}
    \caption{Performance comparison of the hybrid confirmation tree and decision making with AI advice per data set. The mean values of true positive and true negatives rates (x--axis) of the hybrid confirmation tree (blue), humans with AI advice (red), and humans without AI advice (orange) across data sets. Values in blue boxes show the accuracy improvement of the hybrid confirmation tree over humans with AI advice.}
    \label{figs_tpr_fpr}
\end{figure*}

\newpage

\begin{figure}[!t]
    \centering
    \includegraphics[width = 0.8\textwidth]{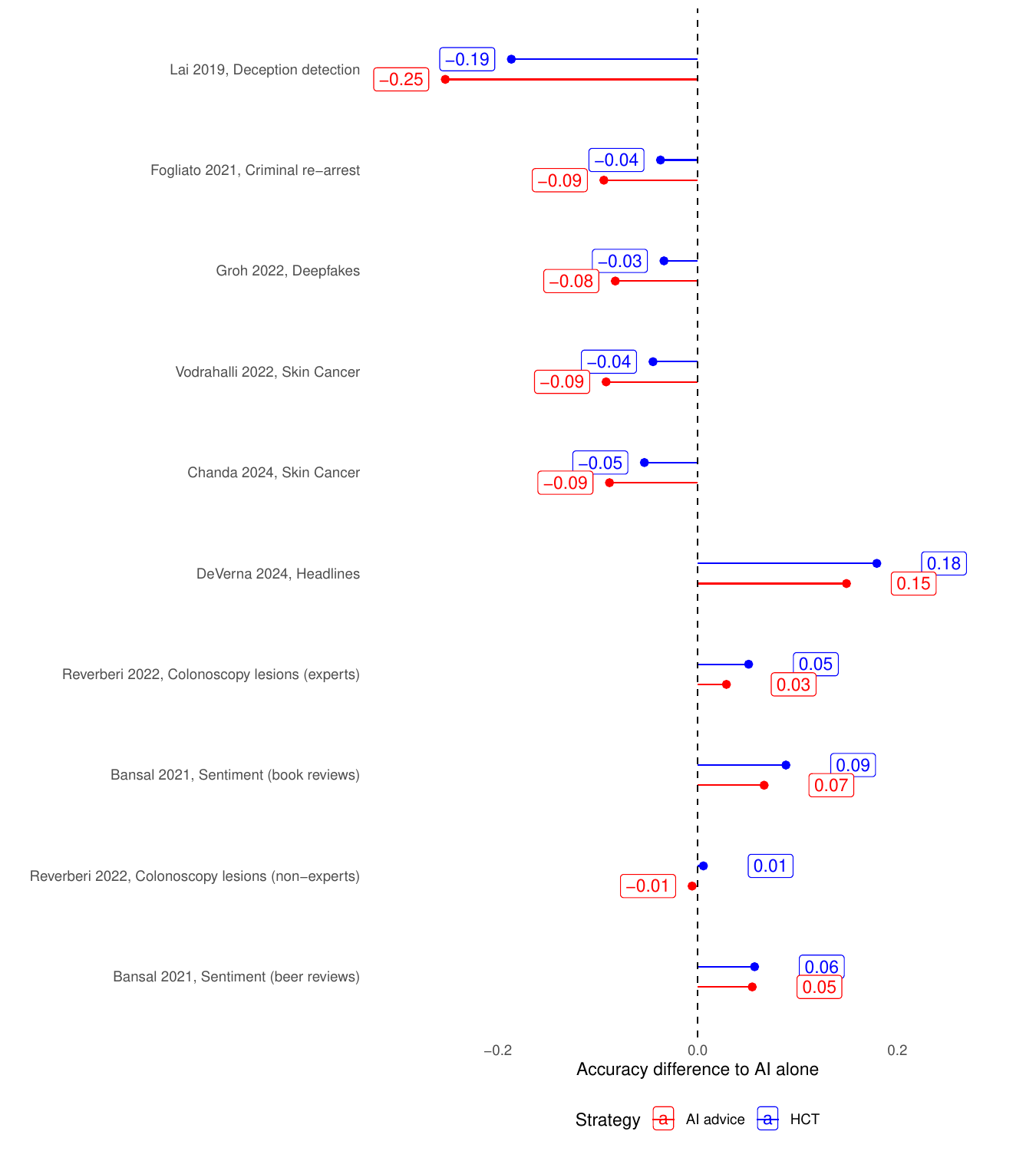}
    \caption{The accuracy of the hybrid confirmation tree (blue) and humans with AI advice (red) against the accuracy of the AI alone (x-axis). Positive (/negative) values indicate the method performs better (/worse) than the AI alone.}
        \label{complementarity}
\end{figure}


\newpage

\newpage

\begin{figure*}[!htbp]
    \centering
    \includegraphics[width = \textwidth]{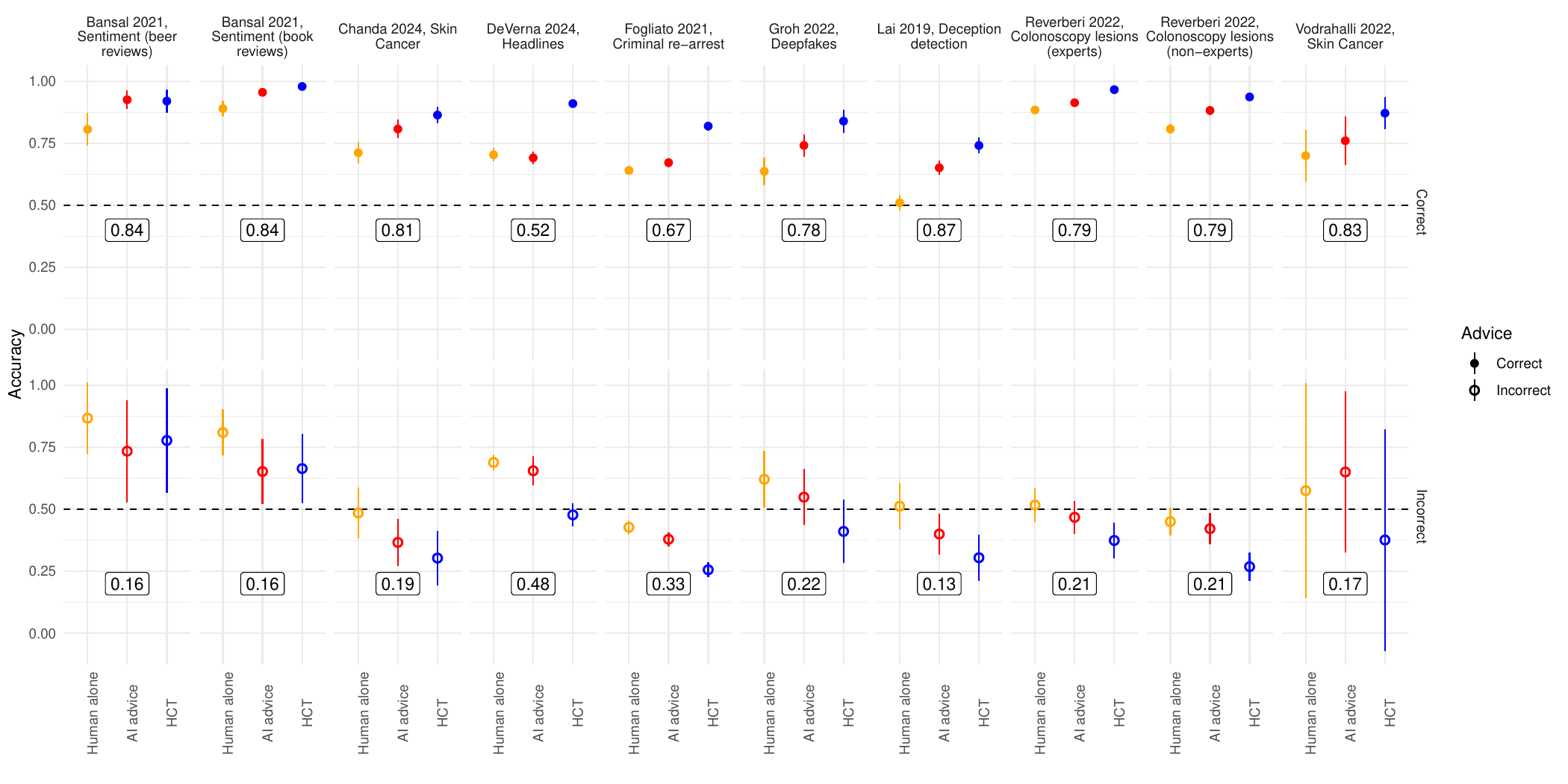}
    \caption{The accuracy of the hybrid confirmation tree (blue), humans with AI advice (red) and humans on their own (orange) for cases where the AI was correct (closed circles) and incorrect (open circles). Results are means across all data sets, the error bars correspond to the 95$\%$ Confidence Intervals (CI). Numbers show frequencies of correct and incorrect AI decisions.}
      \label{figs_domain_correct}
\end{figure*}

\newpage

\begin{figure*}[!htbp]
    \centering
    \includegraphics[width = \textwidth]{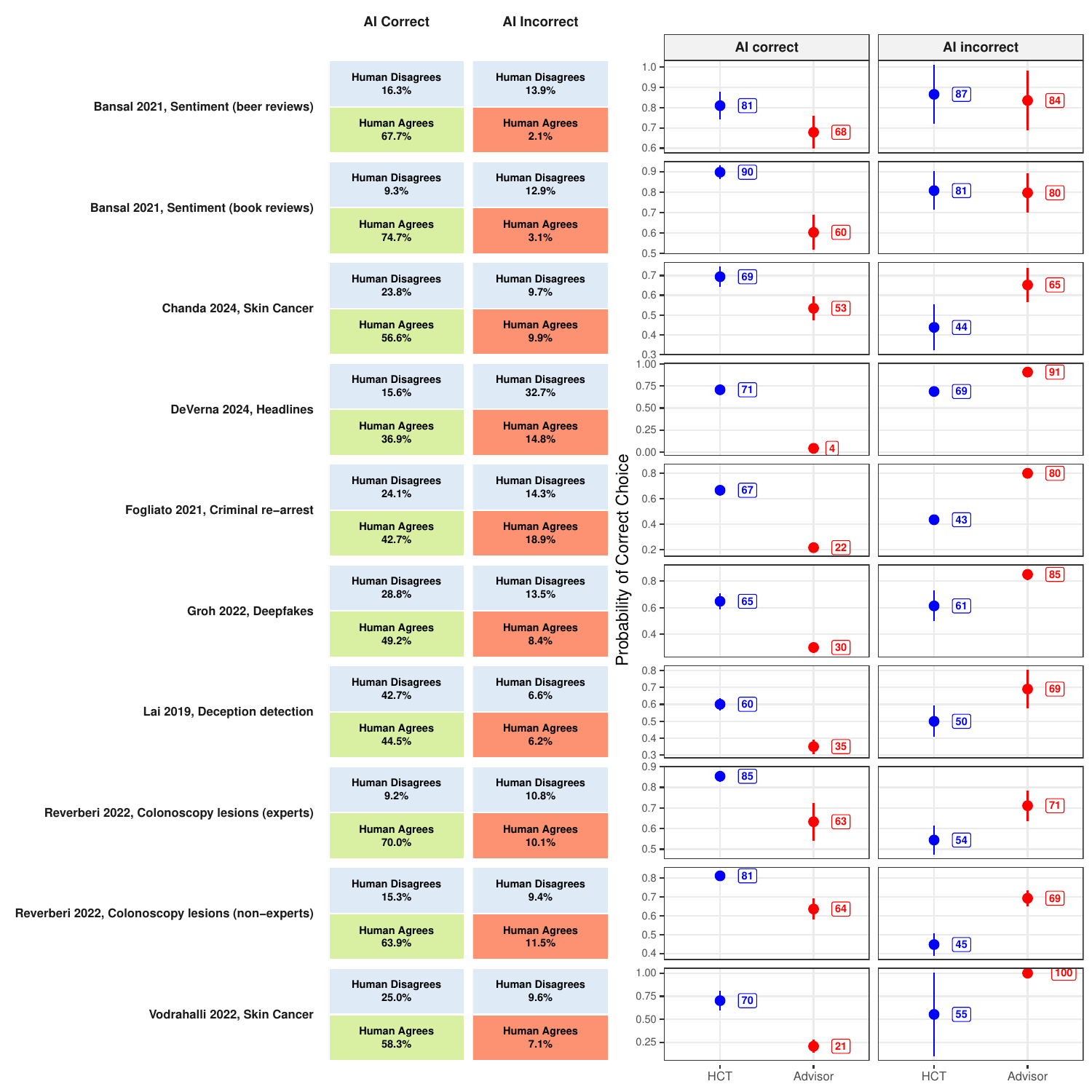}
    \caption{Human-AI agreement and disagreement for correct and incorrect choices for all data sets. Dots-and-whisker plot shows the performance of the hybrid confirmation tree and humans with AI advice for cases where the AI was correct (left) or incorrect (right) and there was human--AI disagreement, meaning either the tie-breaker had to make an independent decision or the human had to revise their judgment in light of AI advice. Results are averages (presented in numbers) within data sets, the error bars correspond to the 95$\%$ CI.}
      \label{figs_domain_correct}
\end{figure*}

\newpage

\begin{figure*}[!htbp]
    \centering
    \includegraphics[width = \textwidth]{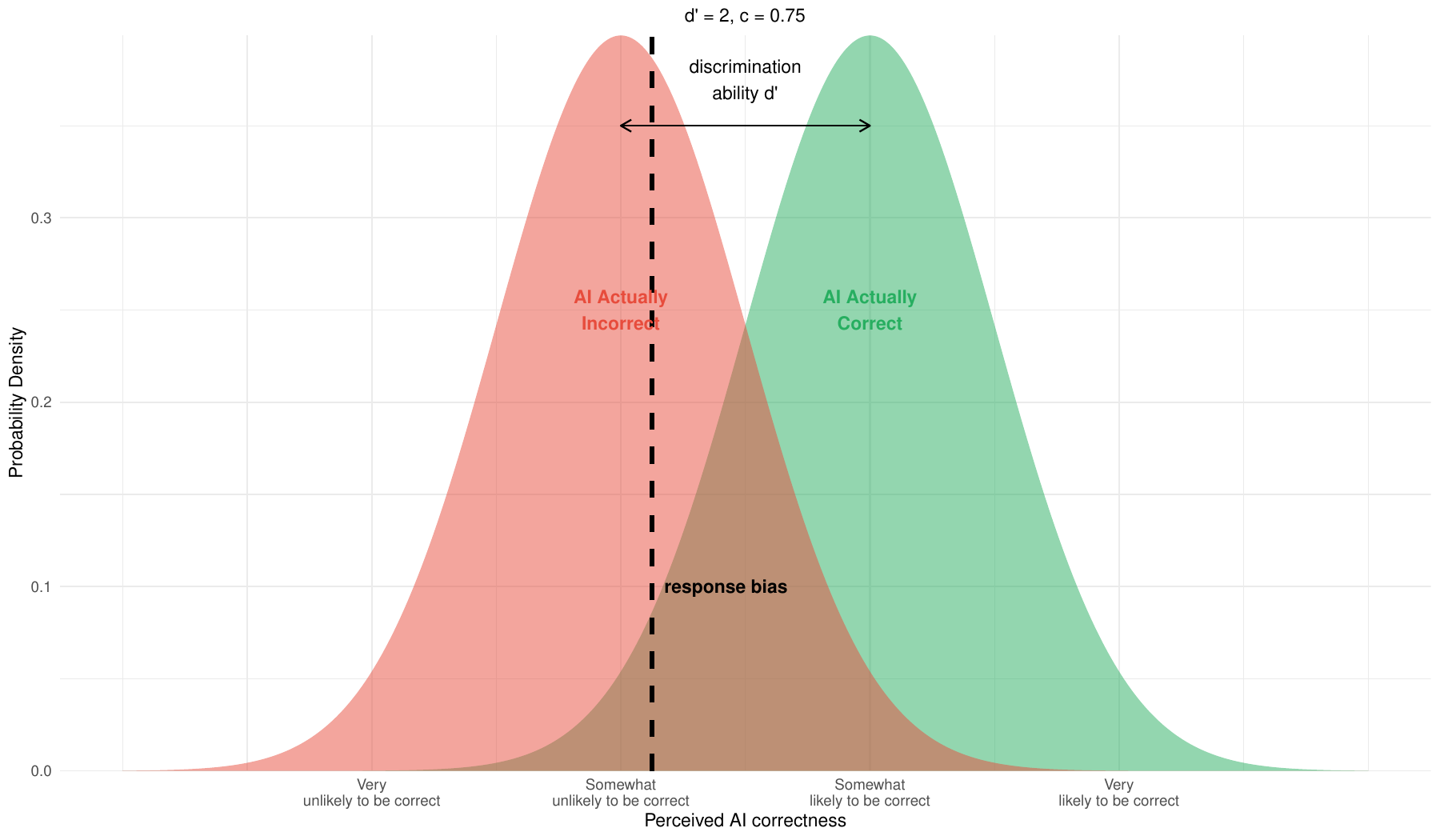}
    \caption{The concept of signal detection theory (SDT) applied to AI advice taking. We conceptualize the AI advice taking situation similar to a typical SDT task where there are two distributions with some overlap between the signal distribution (AI correct, green) and the noise distribution (AI incorrect, red). Depending on the perceived AI advice correctness for a task and the decision maker's response bias (dashed line) people will accept more or less good and bad advice. The decision maker's ability to differentiate correct from incorrect AI advice is measured by the distance between the means of the signal and noise distribution (black arrow on top).}
      \label{sdt_plot}
\end{figure*}

\newpage

\begin{figure*}[!htbp]
    \centering
    \includegraphics[width = \textwidth]{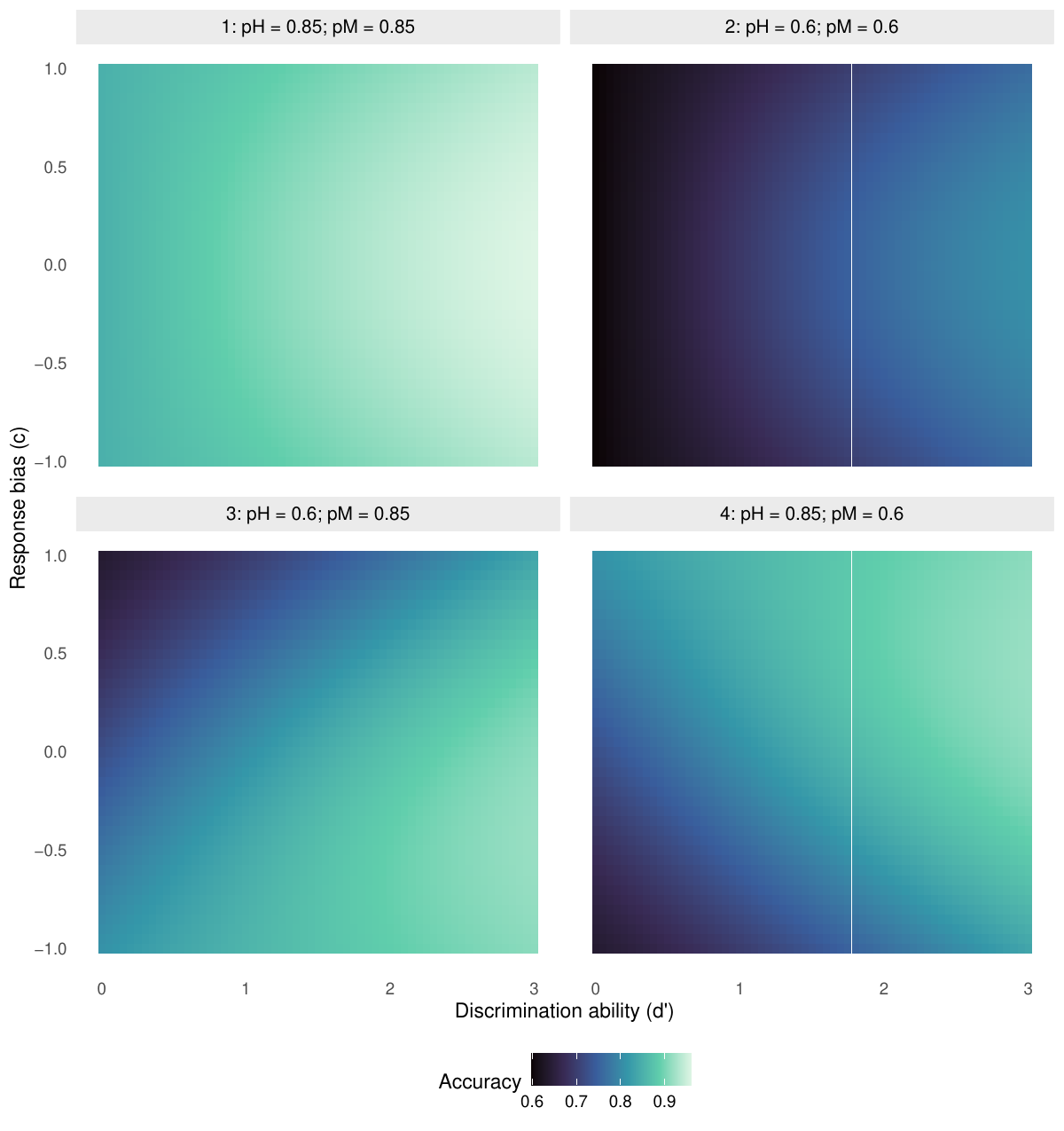}
    \caption{The effect of discrimination ability (x--axis) and response bias (y--axis) on the performance of humans with AI advice (color) as a function of human (pH) and AI accuracy (pM). At equal human and machine accuracy (top--row), the AI advice taking accuracy is maximized at a neutral response bias. When the AI is much better than humans (bottom--left), taking more advice (lower response bias) is beneficial. When the AI is worse than humans (bottom--right), rejecting more advice is beneficial (i.e., higher response bias). Moreover, increasing discrimination ability (as well as greater human--only and/or AI--only accuracies) lead to higher accuracy.}
      \label{sdt_examples}
\end{figure*}

\newpage

\begin{figure*}[!t]
    \centering
    \includegraphics[width = 0.8\textwidth]{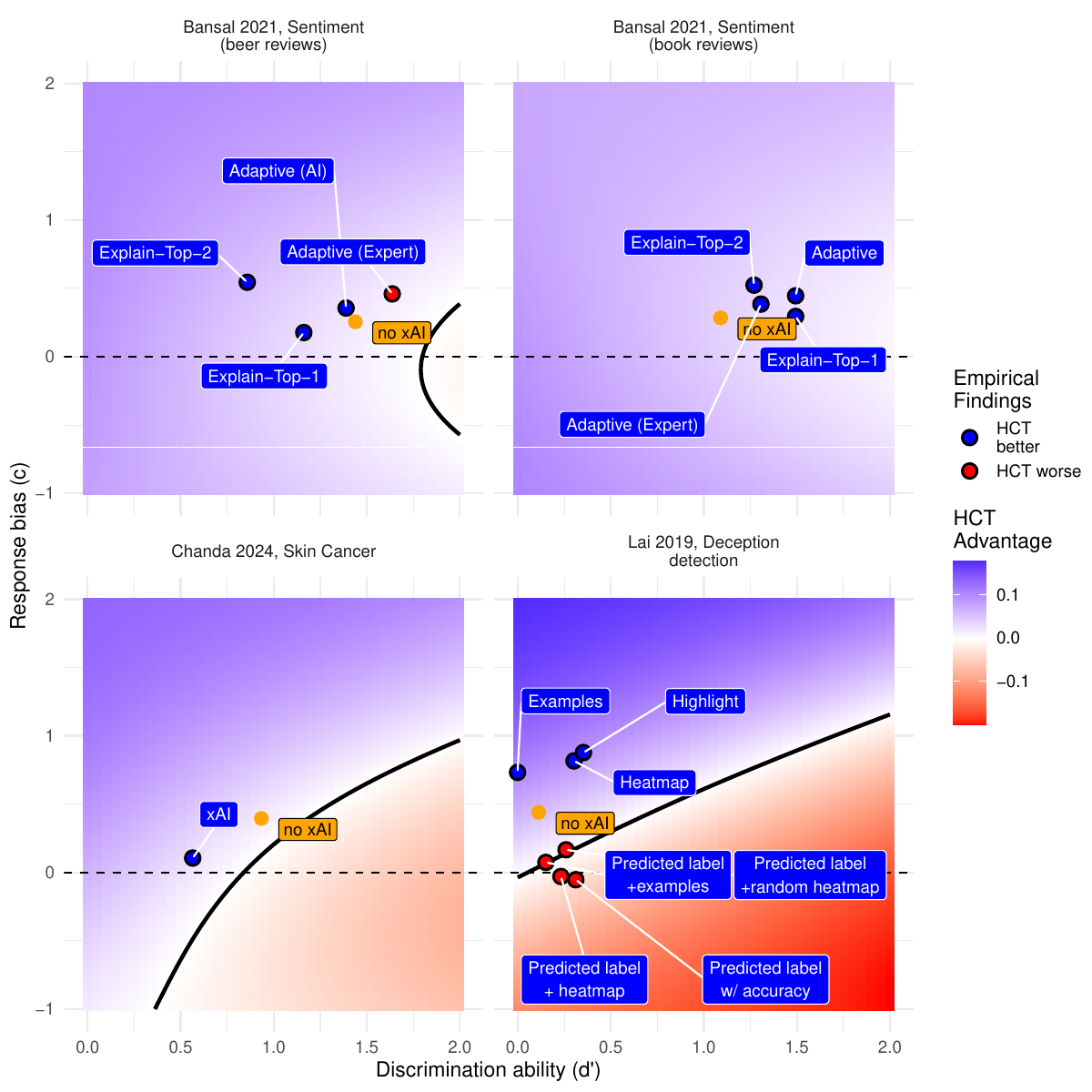}
    \caption{Theoretical comparison of humans with explainable AI advice and the hybrid confirmation tree (HCT) across data sets. Blue (red) background colors indicate that the HCT performs better (worse) than the explainable AI advice. This is shown as a function of the human discrimination ability between correct and incorrect AI advice (x--axis) and the human propensity to rely on AI advice expressed as the response bias, with higher values indicating higher conservatism towards not taking advice (y--axis). These color values are derived analytically. The color fill of the dots indicates the empirical finding of whether the HCT performed better (blue) or worse (red) than humans with explainable AI advice. The orange labels shows the results of the regular AI advice condition (i.e., not explainable AI) as a benchmark.}
    \label{sdt_xai}
\end{figure*}

\newpage

\newpage

\begin{figure*}[!t]
    \centering
    \includegraphics[width = 0.8\textwidth]{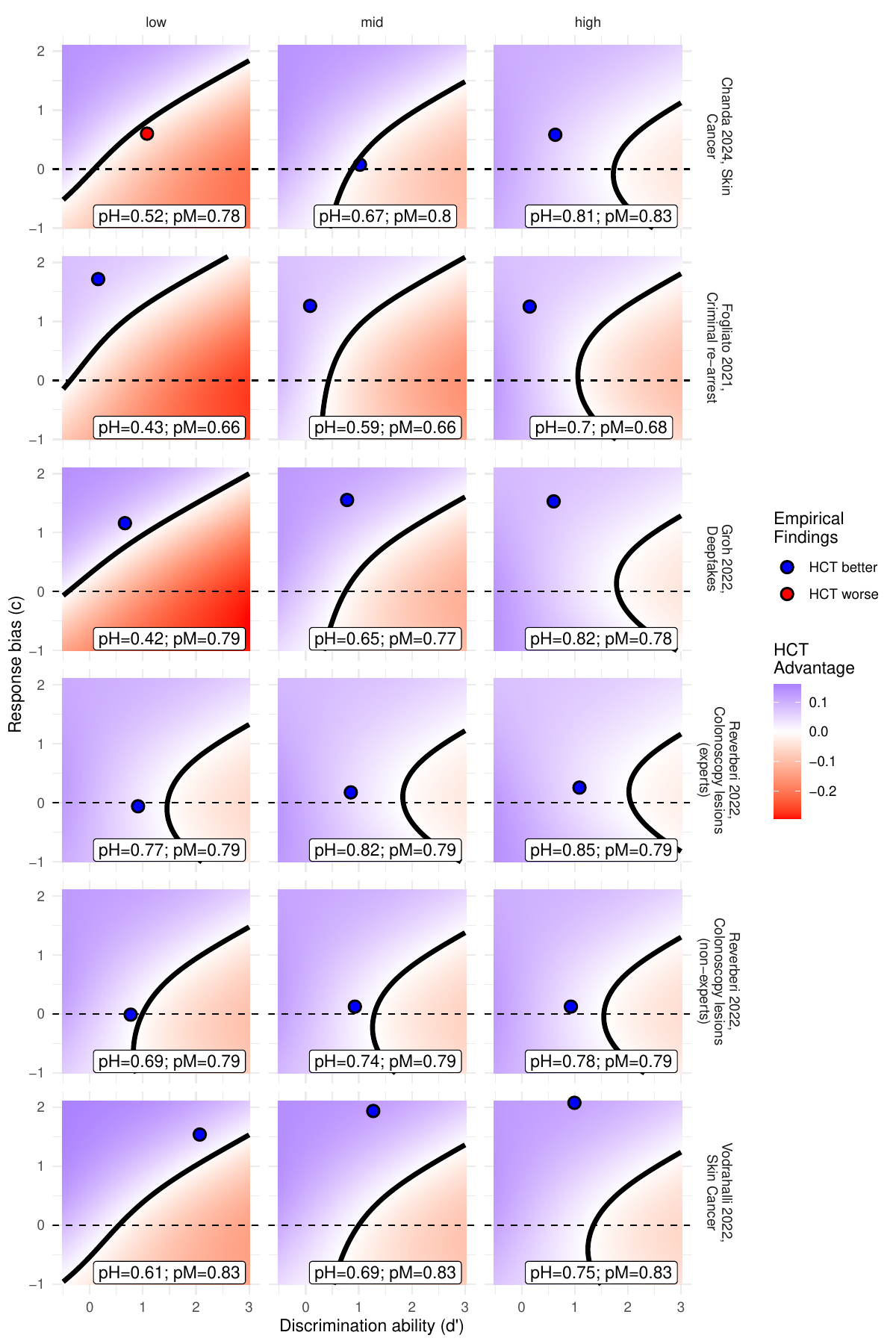}
    \caption{Theoretical comparison of humans with AI advice and the hybrid confirmation tree across data sets and different performance ranks. Blue (red) background colors indicate that the HCT performs better(worse) than the AI--as--advisor. This is shown as a function of the human discrimination ability between correct and incorrect AI advice (x--axis) and the human propensity to rely on AI advice expressed as the response bias, with higher values indicating higher conservatism towards not taking advice (y--axis). These color values are derived analytically. We calculate the average discrimination ability and response bias of the data sets given different levels of human-only performance and locate them in the parameter space (black dots).}
    \label{sdt_xai}
\end{figure*}

\end{document}